\documentclass[fleqn,11pt]{wlscirep}

\usepackage[utf8]{inputenc}
\usepackage[T1]{fontenc}
\usepackage{graphicx}
\usepackage{xcolor}
\usepackage{unicode}
\usepackage{caption}
\usepackage{float}

\newcommand{\Np}{N_{\rm p}}
\newcommand{\Npt}{N_{{\rm p},t}}
\newcommand{\Nph}{\widehat{N}_{\rm p}}
\newcommand{\Npht}{\widehat N_{{\rm p},t}}
\newcommand{\kapht}{\widehat \kappa_{t}}
\newcommand{\kapt}{\kappa_{t}}

\DeclareMathOperator*{\argmax}{arg\,max}

\title{Identifying the temporal dynamics of densification and sparsification in human contact networks}
\author[1]{Shaunette T. Ferguson}
\author[2,3,*]{Teruyoshi Kobayashi}
\affil[1]{Graduate School of Economics, Kobe University, Kobe, Japan}
\affil[2]{Department of Economics, Kobe University, Kobe, Japan}
\affil[3]{Center for Computational Social Science, Kobe University, Kobe, Japan}

\affil[*]{kobayashi@econ.kobe-u.ac.jp}

\begin{abstract}
Temporal social networks of human interactions are preponderant in understanding the fundamental patterns of human behavior. 
In these networks, interactions occur locally between individuals (i.e., nodes) who connect with each other at different times, culminating into a complex system-wide web that has a dynamic composition. 
Dynamic behavior in networks occurs not only locally but also at the global level, as systems expand or shrink due either to: changes in the size of node population or variations in the chance of a connection between two nodes.
Here, we propose a numerical maximum-likelihood method to estimate population size and the probability of two nodes connecting at any given point in time. An advantage of the method is that it relies only on aggregate quantities, which are easy to access and free from privacy issues.
Our approach enables us to identify the simultaneous (rather than the asynchronous) contribution of each mechanism in the densification and sparsification of human contacts, providing a better understanding of how humans collectively construct and deconstruct social networks. 
\end{abstract}
\begin{document}

\flushbottom
\maketitle

\thispagestyle{empty}
\section{Introduction}

Individuals are interacting in unprecedented ways due to advancements in communication technology, which has granted access to human contact data in a variety of social contexts (e.g., mobile calls~\cite{JoKarsai2012NewJPhys,Onnela2007PNAS,kovanen2010,Schalfer2014,Ghosh2019}, texts~\cite{Opsahl2008PhysRevLett,Panzarasa09JASIST}, email~\cite{Klimt2004enron}, face-to-face~\cite{Isella2011JTB,Starnini2013PRL,Barrat2013,Genois2015,kobayashi2019structured}). Our understanding of fundamental human behavioral patterns have benefited considerably from these rich data sources in which individuals (i.e., nodes) establish and break existing connections (i.e., edges) with each other, thus driving the evolution of a complex network structure. To capture the dynamics of these systems in which the contacts between nodes occur intermittently, social networks are often modeled using a temporal representation~\cite{Holme2012PhysRep,Holme2015EPJB}. 

In social systems, contacts tend to occur periodically because individuals have a choice on how and when to engage with others; hence, at a given point in time, the number of active nodes ($N$) and the number of edges ($M$) in the system are changing. Furthermore, many empirical networks exhibit a relationship between total edges and network size that is consistent with a densification scaling property~\cite{Leskovec2005HepAS,Leskovec2007CA_Full,kobayashi2018social}: $M \propto N^{\gamma}$ with $\gamma >1$, in which aggregate edges increase superlinearly in network size. In temporal social networks, this dynamical property between $N$ and $M$ is influenced either by i) fluctuations in population size~\cite{Leskovec2007CA_Full,kobayashi2020densification,kobayashi2021switching}, ii) changing probability of node connection~\cite{kobayashi2020densification}, or iii) both~\cite{kobayashi2021switching}. Given a fixed connection probability and changing size of population, the conventional superlinear scaling emerges i.e., $M \propto N^{\gamma}$ with $\gamma >1$~\cite{kobayashi2020densification}. Conversely, for constant population size and varying connection probability, $M$ exhibits an accelerating growth pattern~\cite{kobayashi2020densification}.

However, many human contact networks exhibit a dynamical $N$-$M$ relationship that is a mixture of the two behaviors, each appearing either as a growth in $M$ along a straight line or an increasing $M$ along an upward sloping trajectory on log-log scale~\cite{kobayashi2020densification,kobayashi2021switching}.
This type of mixed densification scaling usually appears when individuals are free to enter and exit the system, and opportunities to connect are clearly defined (e.g., during lunch in a work setting) or activities are strictly regulated by a schedule (e.g., events at a conference).  
At a conference, for instance, it is expected that attendees will limit socialization during times designated for a keynote talk because they are attentive to the speaker. During coffee break, in contrast, they are free to interact with others.
The emergence of a mixed scaling relationship in temporal social networks suggests that the mechanism that describes the dynamical growth of $M$ in $N$ may be alternating occasionally~\cite{kobayashi2021switching}. 
From this standpoint, a Markov regime-switching model~\cite{Hamilton1994book,hamilton2010regime} is employed in a previous study to estimate the probability that the dynamical source of densification and sparsification is attributed either to changing population size \emph{or} fluctuating intensity in activity level at a given time~\cite{kobayashi2021switching}. 

Here, we develop an alternative approach to identify the extent to which changing population and connection probability concurrently influence the dynamics of densification and sparsification in human contact networks. The proposed method, based on numerical likelihood functions, enables the simultaneous estimation of population size ($= \#$~active nodes $+ ~\#$~isolated nodes) and connection probability in different social networks using a series of ($N,M$) observations, each corresponding to a given temporal snapshot. 
By taking this approach, we can gain insight not only into the independent contribution of the two mechanisms but also into how their co-movement influences the emergence of a mixed scaling. 
While contact lists (or event sequences) usually allow us to observe the number of active individuals who made at least one contact, the number of inactive individuals who were present but have never interacted (i.e., isolated nodes) are often unknown. Our approach also provides an estimate for the number of isolated nodes by relying only on the total numbers of active nodes and edges at a given point in time.

\section{Methods}

\subsection{Data}

We use the following four temporal human-contact networks collected by the SocioPatterns collaboration~\cite{Sociopatterns}:
\begin{itemize}
\item \textbf{Hospital~\cite{Vanhems:2013}}: Contacts between patients, nurses and doctors at a hospital in Lyon, France on December 7, 2010.
\item \textbf{Workplace~\cite{Genois:2017}}: Contacts between employees at an office building in France on June 27, 2015.
\item \textbf{IC2S2-17~\cite{Genois2019}}: Contacts between conference attendees at the International Conference on Computational Social Science 2017 at GESIS in Cologne, Germany on July 11, 2017.
\item \textbf{WS-16~\cite{Genois2019}}: Contacts between participants at the Computational Social Science Winter Symposium 2016 at GESIS in Cologne, Germany on December 1, 2016.
\end{itemize}
For each data set, interaction between individuals occurs in a physical location, and Radio Frequency Identification (RFID) sensors detect a contact when one person is within 1.5 meters of another~\cite{Genois2019,Vanhems:2013,Genois:2017}. 
Contacts are recorded at 20-second intervals. 
Such high-resolution data have been frequently used to discover temporal patterns in human behavior~\cite{Cattuto2010PlosOne,Holme2012PhysRep,masuda2016guide,karsai2018bursty} or to explore how infectious diseases spread through human contacts~\cite{salathe2010high,Stehle2011PLOS,masuda2017temporal}.

\begin{figure}[t]
    \centering
    \includegraphics[width = 0.9\textwidth]{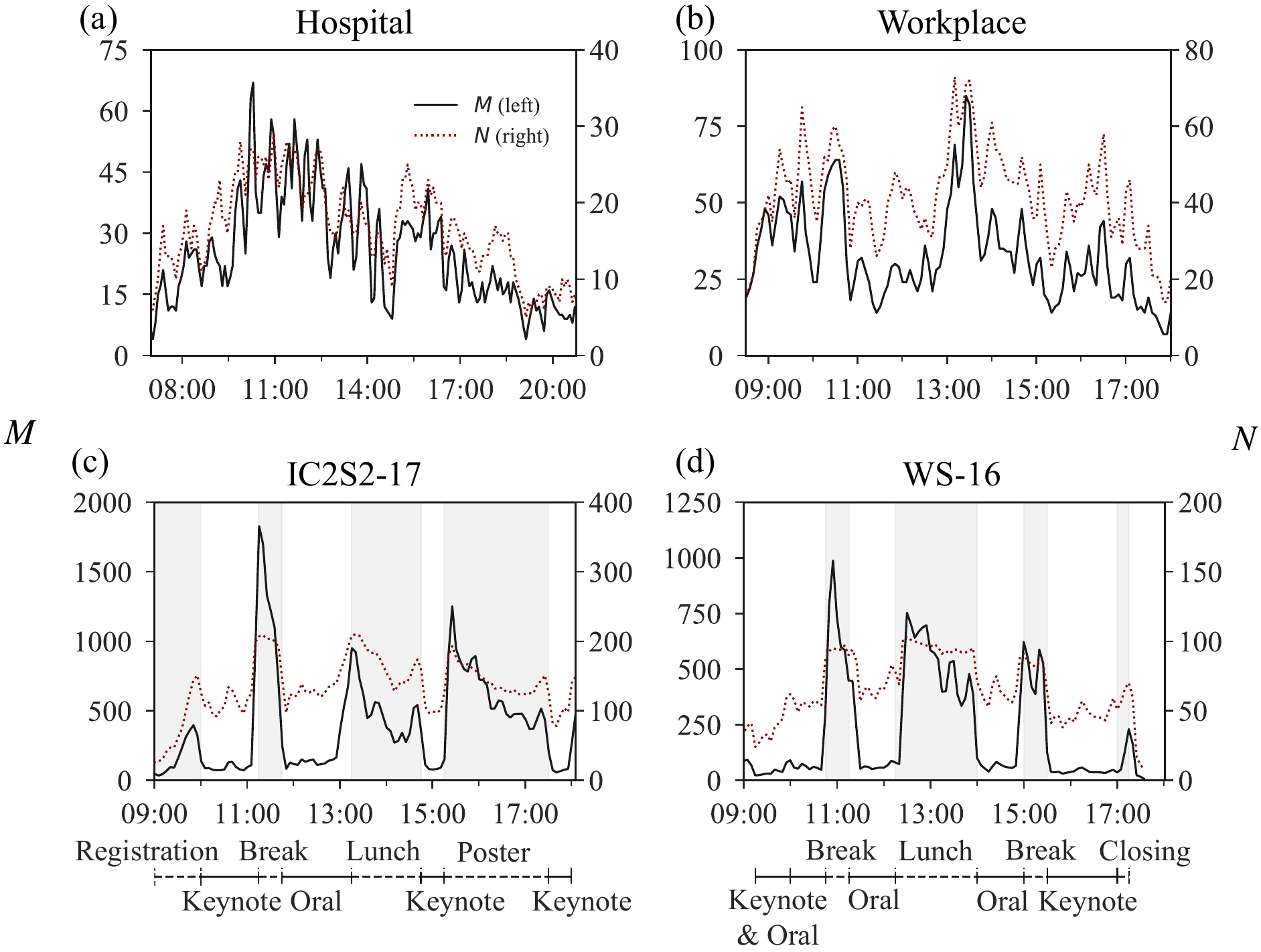}
    \caption{Evolution of number of edges $M$ and active nodes $N$ in face-to-face networks. The following days are shown for each data set: (a) Hospital on December 7, 2010, (b) Workplace on June 27, 2015, (c) IC2S2-17 on July 11, 2017 and (d) WS-16 on December 1, 2016. Timeline below panels c and d identify time windows for scheduled events. Gray shading highlights unrestricted sessions i.e., registration, break, lunch, poster session and closing remarks.}
    \label{fig:MNline0}
\end{figure}

We take advantage of the time-resolved data to explore the temporal dynamics of densification and sparsification in the data sets, by converting them to temporal networks with unweighted and undirected edges. We segment a data set into a snapshot sequence  (i.e., a series of networks that are ordered in time~\cite{Cazabet2019}), which we construct as sliding time windows. A time window has a duration of 10 minutes and consecutive windows have a 5-minute overlap between them. Then, we connect two nodes if they have at least one contact within the time window, and we extend this to all other time windows to obtain a sequence of snapshots.
A node is considered to be active if we detect that it is involved in one or more contact events for a given network snapshot. 
The numbers of active nodes and edges in a snapshot are denoted by $N$ and $M$, respectively.
The observed $N$ and $M$ are shown in Fig.~\ref{fig:MNline0} (See Fig.~\ref{fig:MNline1} in Supplementary Information for different days).

\subsection{Estimation}

\subsubsection{Dynamic hidden-variable model}
To explore the densification and sparsification dynamics in temporal networks, we employ a hidden-variable (or a fitness) model with a temporal dimension~\cite{Caldarelli2002PhysRevLett,Boguna2003PhysRevE,kobayashi2020densification,kobayashi2021switching}. The probability that two nodes $i$ and $j$ are connected in time interval $[t,t+\Delta t]$ (henceforth, we refer to as time interval $t$) is given by
\begin{align}
p_{ij,t} = 1-e^{-\kappa_{t} a_{i} a_{j}},\quad{i, j = 1,\ldots,\Npt,\; t = 1,\ldots,T},\label{eq:probfn}
\end{align}
where $a_{i}$ is node $i$'s intrinsic activity level and is assumed to be uniformly distributed on $[0,1]$. 
Note that the dynamical source of networks is decomposed into two factors: $\Npt$ and $\kapt$. In time interval $t$, the overall activity of nodes is captured by $\kappa_{t}>0$, which encapsulates changing  activity levels due to prespecified schedule, circadian rhythm, etc., while the total number of nodes (i.e., combined sum of active and inactive nodes) is denoted by $\Npt$. 
It should be noted that the number of active individuals $N$, at a given time, can be directly observed from contact lists, but the potential number of individuals (i.e., population) in a system is not usually known because contact events naturally exclude non-interacting individuals. 
Due to the lack of information on population, it is generally not obvious to what extent variations in $N$ and $M$ could be explained by changes in population or activity. Our model takes into account the two possible factors, population and overall activity level, in explaining the observed behaviors of $N$ and $M$, which cause densification and sparsification of temporal networks.

As an alternative to the connecting probability in Eq.~\eqref{eq:probfn}, we also show the results for the following specification: 
\begin{align}
p_{ij,t} = \kappa_{t} a_i a_j,\quad{i, j = 1,\ldots,\Npt, t = 1,\ldots,T}.\label{eq:probfn0}
\end{align}
This specification is employed in previous studies~\cite{kobayashi2020densification,kobayashi2021switching}, and we confirm that the essential results do not change compared to the baseline model based on Eq.~\eqref{eq:probfn}.

\subsubsection{Numerical maximum-likelihood estimation}

We estimate the parameters $(\kapt,\Npt)$ for a given $(N_t,M_t)$ in time interval $t$, using a numerical maximum-likelihood method. 
Let $\Theta_\kappa\equiv \left\{\kappa^{(1)},\ldots,\kappa^{(L_{\kappa})}\right\}$ and $\Theta_{\rm p}\equiv \left\{\Np^{(1)},\ldots,\Np^{(L_{\rm p})}\right\}$ be the sets of all possible values for $\kappa$ and $\Np$, respectively. The Cartesian product of two sets $\Theta_\kappa$ and $\Theta_{\rm p}$ is given as
\begin{align}
   \Theta =\left\{(\kappa,\Np)|\kappa\in\Theta_\kappa, \Np\in\Theta_{\rm p} \right\}.
\end{align}
We define $\bm{\theta}^\ell\in\Theta$ as the $\ell$-th element of the set $\Theta$ for $\ell=1,\ldots,|\Theta|$, where $|\Theta|= L_\kappa L_{\rm p}$ is the cardinality of $\Theta$, i.e., the total number of combinations $(\kappa,\Np)$.

\begin{figure}[t]
    \centering
    \includegraphics[scale=0.65]{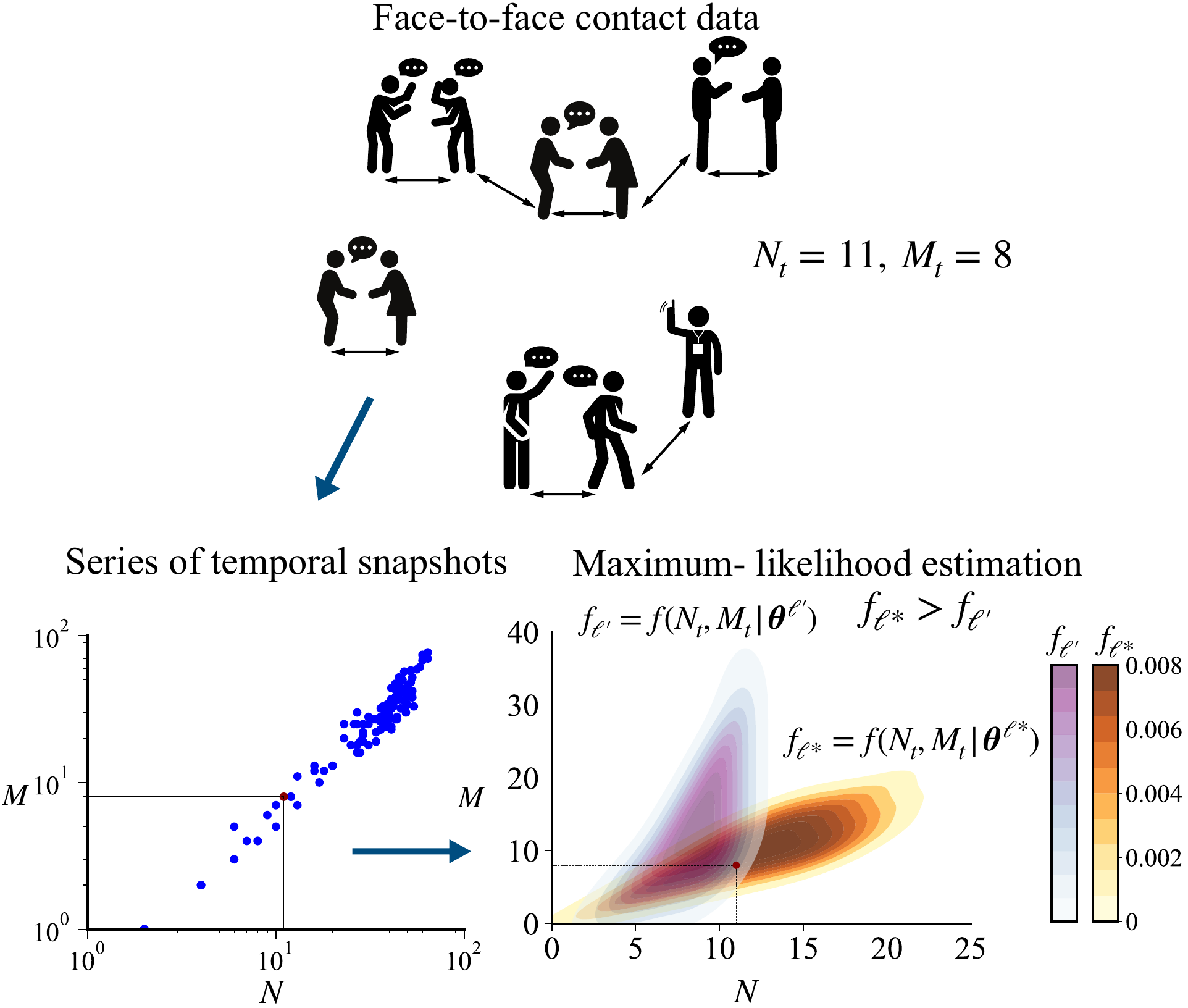}
    \caption{Schematic of maximum-likelihood estimation of $\kappa$ and $\Np$. The top panel shows contact data that gives a combination $(N_t,M_t)$. The sequence $\{(N_t,M_t)\}_{t=1}^T$ is plotted in the $N$-$M$ space, in which a particular combination of $(11,8)$ is highlighted in red (bottom left). The joint distributions of $(N,M)$, or likelihood functions, are generated using the hidden-variable model for different combinations of $(\kappa,\Np)=\bm{\theta}$, with each indexed by $\ell^{\prime}$ and $\ell^{\ast}$ (bottom right). A likelihood function gives the probability of observing a network with $N$ nodes and $M$ edges, for a given combination of $(\kappa,\Np)$. The maximum-likelihood estimators, denoted by $\kapht$ and $\Npht$, are given by a combination of $\kappa$ and $\Np$ associated with the maximum-likelihood function $f_{\ell^\ast}=f(N_t,M_t|{\bm{\theta}}^{\ell^{\ast}})$.}
    \label{fig:MLEschem}
\end{figure}

Our maximum-likelihood estimation proceeds as follows:
\begin{enumerate}
    \item For a given $\bm{\theta}^\ell$, generate an unweighted and undirected network based on probabilities $\{p_{ij}\}$ for $i>j$. By repeating the network generation $S$ times, one can obtain a sequence of combinations $\left\{\left(N^{(s)},M^{(s)}\right)\right\}_{
    s=1}^{S}$,  
    where $N^{(s)}$ and $M^{(s)}$ respectively denote the number of active nodes and the number of edges observed in the $s$-th simulation. We set $S=10^4$.
    
    \item Count the number of appearances of each unique combination in $\left\{\left(N^{(s)},M^{(s)}\right)\right\}_{
    s=1}^{S}$ and express as a fraction of the number of runs $S$ to get the joint distribution $f_\ell(N, M|\bm{\theta}^\ell)$, i.e., the likelihood function for a given $\bm{\theta}^\ell$.
    
    \item Repeat steps 1 and 2 to obtain a set of likelihood functions $\left\{f_\ell(N, M|\bm{\theta}^\ell) \right\}_{\ell=1}^{|\Theta|}$.

    \item Select $\ell=\ell^*(\leq |\Theta|)$ such that ${f}_{\ell^{*}}(N_t,M_t|\bm{\theta}^{\ell^*})$ yields the highest probability for a given empirical observation $(N_t,M_t)$. 
    The maximum-likelihood estimators $\kapht$ and $\Npht$ are thus given by  
    \begin{align}
    \left(\kapht,\Npht\right) = \bm{\theta}^{\ell^*},
    \end{align}
    where $\ell^* =\argmax_{\ell}\: f_\ell(N_t,M_t|\bm{\theta}^{\ell})$.

    \item Repeat steps 1--4 for all time intervals $t = 1,\ldots, T$.
\end{enumerate}
A schematic of the estimation method is presented in Fig.~\ref{fig:MLEschem}.

\subsection{Validation Analysis}

\begin{figure}[t]
    \centering
    \includegraphics[scale=0.7]{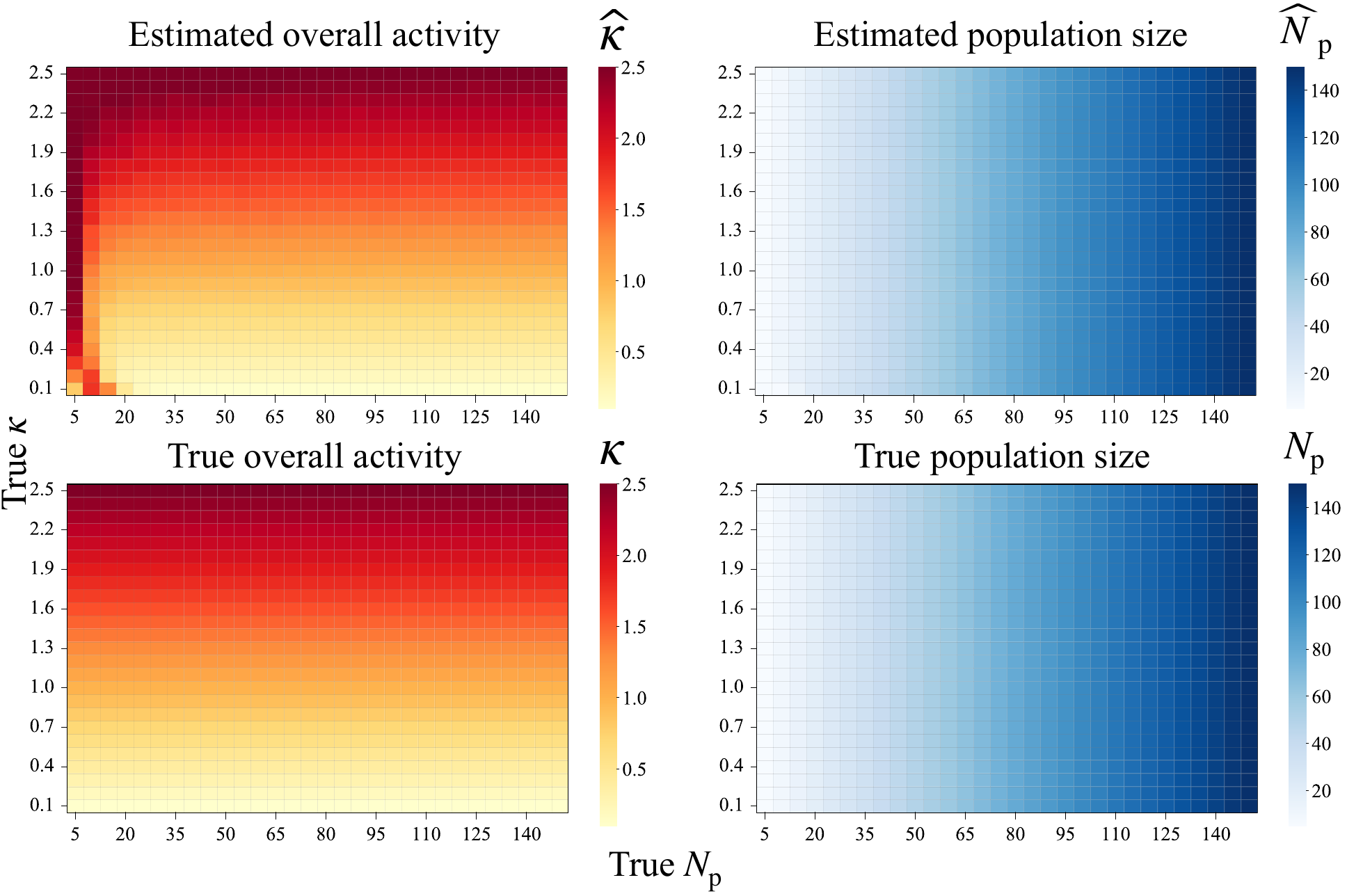}
    \caption{Validation of the maximum-likelihood estimation method. The upper panels show estimated overall activity level $\widehat{\kappa}$ and population size $\Nph$, given the respective true values of $\Np$ ranging from 5 to 150 (incremented by 5) and $\kappa$ ranging from 0.1 to 2.5 (incremented by 0.1). The performance of the maximum-likelihood estimators are respectively assessed against the true overall activity $\kappa$ and true population size $\Np$ shown in the lower panels. The estimates $\Nph$ and $\widehat{\kappa}$ are obtained based on Eq.~\eqref{eq:probfn}.}
    \label{fig:validp1}
\end{figure}

We perform a validation analysis to assess the accuracy of our numerical maximum-likelihood method in estimating the model parameters. For each combination of the true values ($\Np$, $\kappa$), we generate synthetic networks based on the baseline model (Eq.~\ref{eq:probfn}) and apply the estimation method to obtain $\Nph$ and $\widehat{\kappa}$. Then we take the average of the respective estimated values over 1,000 runs.

Based on the comparison of estimated values with their respective true values, the maximum-likelihood estimators perform well in recovering the true population size and overall activity (Fig.~\ref{fig:validp1}). It should be noted, however, that $\widehat{\kappa}$ is sensitive to small values of the true population size (i.e., $\Np \le 50$), in that $\widehat{\kappa}$ overestimates true $\kappa$ (Fig.~\ref{fig:validp1}, \emph{left}).
For larger population sizes (e.g., $\Np> 50$), the performance of the estimation method improves considerably, with deviations, if any, being much smaller. The reason for the low accuracy when $\Np$ is small is that our method relies on $N$ and $M$ to identify the most likely combination of the model parameters; a particular combination ($N$, $M$) does not necessarily have a one-to-one correspondence with a particular ($\kappa$, $\Np$)-combination especially when the network is small, thereby making it possible to see large deviations as exhibited in Fig.~\ref{fig:validp1} (\emph{left}). We also show another validation in which one of the two parameters is fixed (Fig.~\ref{fig:validp1_alternating}).

\section{Results}
\subsection{Evolution of \texorpdfstring{$\kappa$}{} and \texorpdfstring{$\Np$}{} in temporal social networks}

\begin{figure}[t]
    \centering
    \includegraphics[scale=0.75]{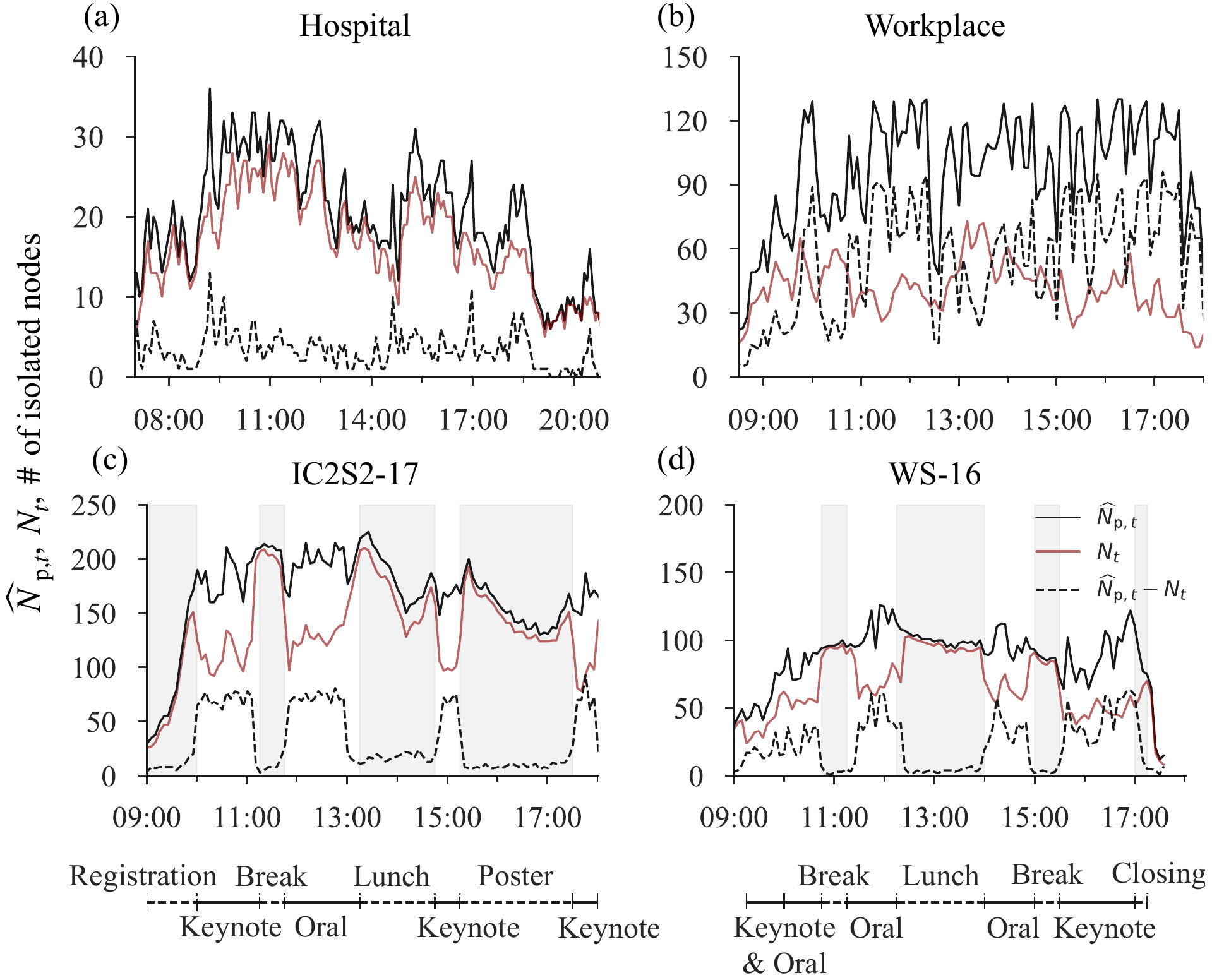}
    \caption{Estimated size of network population, $\Npht$, number of active persons, $N_t$, and total isolated nodes, $\Npht -N_t$ for (a) Hospital, (b) Workplace, (c) IC2S2-17, and (d) WS-16. Timelines at the bottom identify time windows for conference schedule. Gray shading highlights unrestricted sessions i.e., registration, break, lunch, poster session and closing remarks.}
    \label{fig:p1estimates_np}
\end{figure}

\begin{figure}[t]
    \centering
    \includegraphics[scale=0.75]{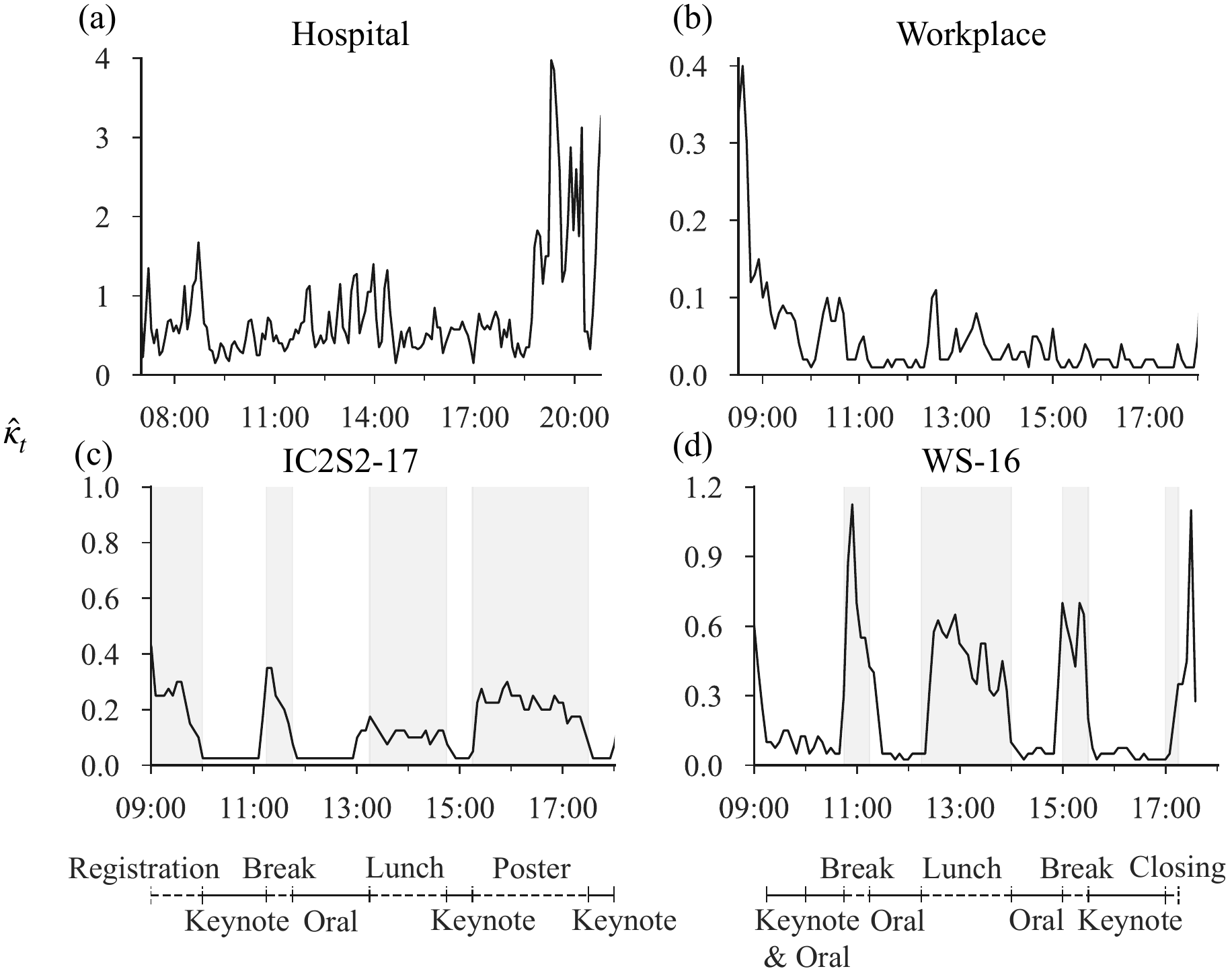}
    \caption{Estimated overall activity level, $\kapht$. (a) Hospital, (b) Workplace, (c) IC2S2-17, and (d) WS-16. Timelines at the bottom identify time windows for conference schedule. Gray shading highlights unrestricted sessions i.e., registration, break, lunch, poster session and closing remarks.}
    \label{fig:p1estimates_kp}
\end{figure}

Estimation results for $\Np$ and $\kappa$ are shown in Figs.~\ref{fig:p1estimates_np} and \ref{fig:p1estimates_kp}, respectively.
Fluctuating $\Npht$ and $\kapht$ in the four data sets indicate that, quite often, both are changing simultaneously. Similar findings are seen in other days for the model based on Eq.~\eqref{eq:probfn} (Figs.~\ref{fig:p1estimates_np1} and \ref{fig:p1estimates_k1}) and also for the alternative probability based on Eq.~\eqref{eq:probfn0} 
(Figs.~\ref{fig:p0estimates_np}--\ref{fig:p0estimates_k1}).  

A source of these shifts in $\kappa$ and $\Np$ would be stemming from situational conditions that may affect human behavior in each location.
One example is a prespecified schedule in an academic conference that rules the behavior of participants~\cite{barrat2010social,barrat2013empirical,kibanov2015web,kobayashi2021switching}. 
For the IC2S2-17 and WS-16 data, we can compare the shifts in the estimated values with the official programs that are available publicly~\cite{IC2S217prog,WS16prog}.
In contrast, a strict schedule of activities is not stipulated in the Hospital and Workplace data, thus precluding a similar kind of assessment.

\subsubsection{Dynamic behavior of estimated population size \texorpdfstring{$\Npht$}{}.}
Sporadic fluctuations of $\Npht$ in Hospital and Workplace (Figs.~\ref{fig:p1estimates_np}a and b) stands in contrast to that of IC2S2-17 and WS-16 (Figs.~\ref{fig:p1estimates_np}c and d), in which $\Npht$ exhibits more systematic variations. Prior to the first keynote talk of IC2S2-17, $\Npht$ increases steadily as expected during a period when participants are arriving at the venue; however, it declines during poster session (Figs.~\ref{fig:p1estimates_np}c). The poster session precedes the final keynote talk; hence the decline in population size (Fig.~\ref{fig:p1estimates_np}c after 15:00) may reflect the exit of participants who, based on the subsequent rise in $\Npht$ shortly after (Fig.~\ref{fig:p1estimates_np}c, 17:00), reconvene for the keynote speech (Fig.~\ref{fig:p1estimates_np}c, 17:30). 
In WS-16, the population size is also high during oral and keynote sessions and a noticeable decline is seen during the closing remarks, which is the final event of the day (Fig.~\ref{fig:p1estimates_np}d, 17:00). In Fig.~\ref{fig:p1estimates_np1}d, WS-16 has a similar schedule to that of IC2S2-17 (Fig.~\ref{fig:p1estimates_np}c) and similar movements in $\Npht$, which grows during registration but subsides during poster session before increasing again prior to the start of the final keynote speech.

In most of the data sets, total active individuals $N_t$ follows closely the population size, which is the maximum possible value of nodes that can be active at a given time (i.e, $N_t \le \Npt$). 
From the estimated population size, we can compute the number of resting nodes as $\Npht-N_t$ (Fig.~\ref{fig:p1estimates_np}, broken line). 
Resting nodes reflect a realistic but generally unobservable feature of dynamic networks, that of isolated individuals who are not in direct contact with any other individual in the system~\cite{kobayashi2020densification}. In conference data, total isolated nodes exhibit a systematic correspondence with activities; few individuals are isolated during registration, break, lunch, and poster session, while elevated levels are seen for keynote talks and oral sessions (Figs.~\ref{fig:p1estimates_np}c and d, broken line). In Hospital data, total isolated nodes is fairly small (close to zero in many instances); however, this is not unnatural in such high-contact environments where hospital staff are frequently engaging each other and/or attending to patients (Fig.~\ref{fig:p1estimates_np}a, broken line). 
In contrast, the number of isolated nodes in Workplace data is generally high, up to three times $N_t$ (Fig.~\ref{fig:p1estimates_np}b).

\subsubsection{Dynamic behavior of estimated overall activity \texorpdfstring{$\kapht$}.}

The estimated activity parameter, $\kapht$, is high during unrestricted sessions at both conferences, signaling intense interactions between participants (Figs.~\ref{fig:p1estimates_kp}c and d, shading). However, during keynote talks and oral sessions, $\kapht$ fluctuates around much smaller values. This suggests that attendees have a greater chance of making contact with each other during registration, coffee break, lunch and poster session than during the oral sessions. Although $\kapht$ declines and remains very low for the duration of keynote talks and oral sessions, our method still detects slight variations, suggesting that $\Npht$ is not the only dynamical parameter at play. 
Fig.~\ref{fig:p0estimates_kp} shows estimated overall activity for the same days based on an alternative probability, Eq.~\eqref{eq:probfn0}.

In contrast, $\kapht$ changes erratically in Hospital and Workplace data. A discernible pattern that corresponds with coordination in movement or activity, as seen in conference data, is not exhibited (Figs.~\ref{fig:p1estimates_kp}a and b). Nevertheless, for Hospital data, $\kapht$ is highest at the end of the day (Fig.~\ref{fig:p1estimates_kp}a) when there is also a diminution in population size (Fig.~\ref{fig:p1estimates_np}a), while for Workplace data, $\kapht$ is highest at the beginning of the day (Fig.~\ref{fig:p1estimates_kp}b) when $\Npht$ is increasing (Fig.~\ref{fig:p1estimates_np}b).  At these times, the behaviors of $N$ and $M$ reflect the dual impact of a sharp rise in $\kapht$ as individuals leave the Hospital network (thereby reducing $\Npht$) or individuals in Workplace join the system (thereby increasing $\Npht$).

\subsection{Time-varying contribution of \texorpdfstring{$\Np$}{} and \texorpdfstring{$\kappa$}{} to the emergence of densification scaling}

\begin{figure}[t]
    \centering
    \includegraphics[scale = 0.75]{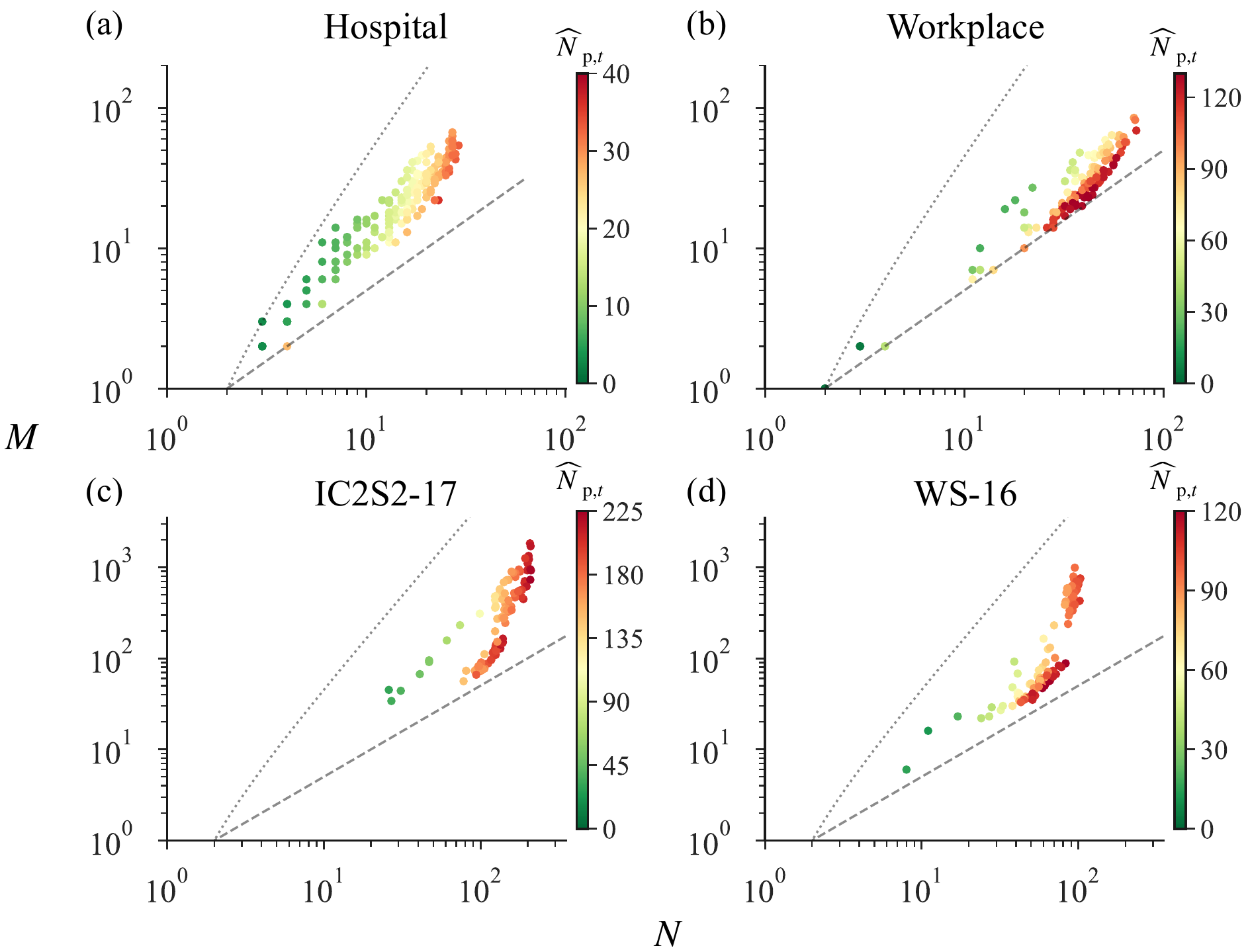}
    \caption{Densification scaling and changes in estimated population size in face-to-face networks. $N$-$M$ scaling plots are shown for (a) Hospital on December 7, 2010 (b) Workplace on June 27, 2015 (c) IC2S2-17 on July 11, 2017 and (d) WS-16 on December 1, 2016. Each dot represents a snapshot of the network and colors denote estimated population size $\Npht$ based on the respective color bar. Gray dashed and dotted lines show theoretical lower ($M = N/2$) and upper ($M = N(N-1)/2$) bounds. Estimates are based on Eq.~\eqref{eq:probfn}.}
    \label{fig:csnp_p1}
\end{figure}

\begin{figure}[t]
    \centering
    \includegraphics[scale = 0.75]{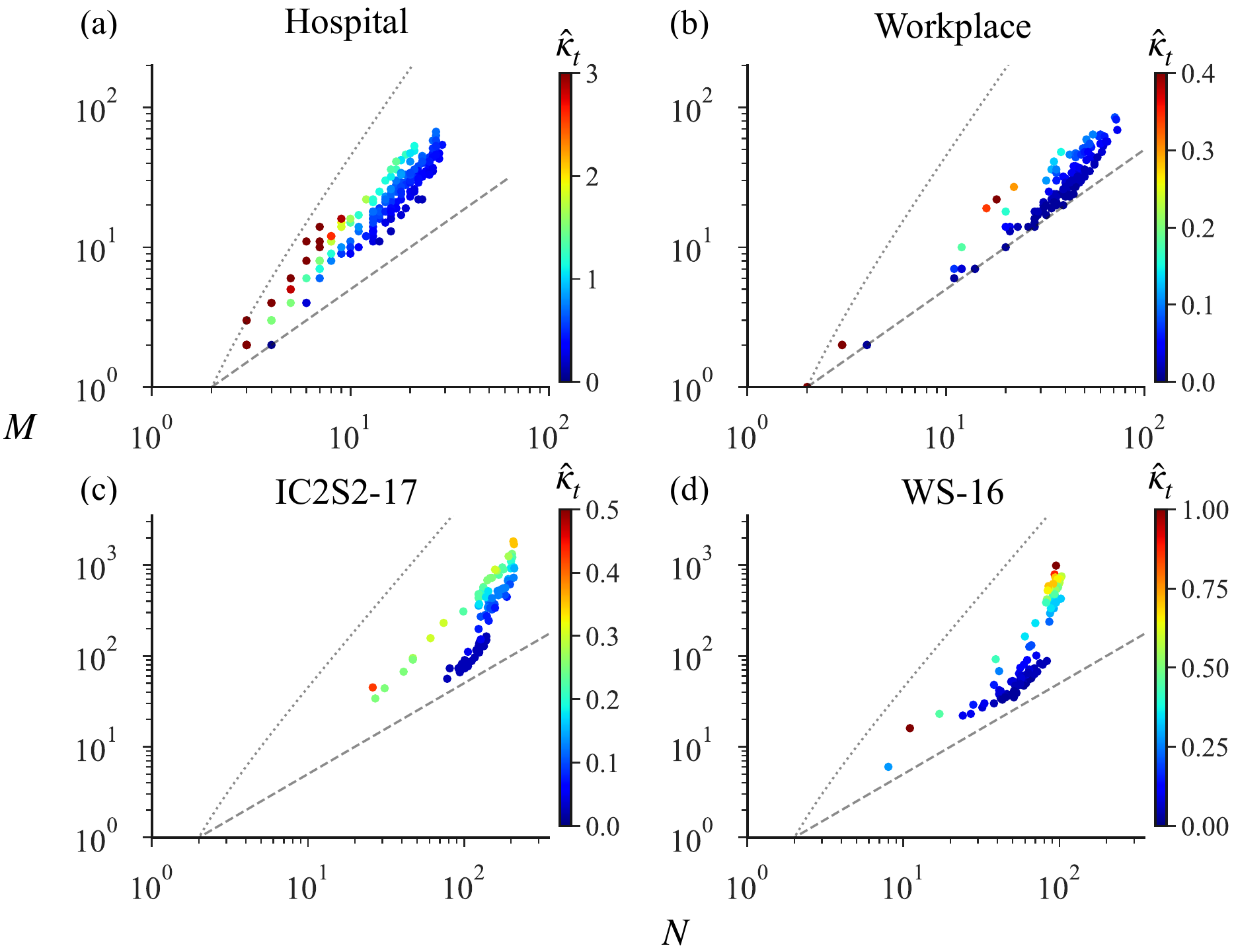}
    \caption{Densification scaling and changes in overall activity $\kapht$ in face-to-face networks. See the caption of Fig.~\ref{fig:csnp_p1} for details.}
    \label{fig:csk_p1}
\end{figure}

We now examine the dynamical relationship between the number of active nodes $N$ and the number of edges $M$ in empirical data to identify the source of densification scaling in social networks.  Figs.~\ref{fig:csnp_p1} and \ref{fig:csk_p1} demonstrate the relationship between $N$ and $M$ based on a series of temporal snapshots for each data set, and the respective color scales denote changing levels of population size $\Npht$ and overall activity $\kapht$. All data sets exhibit a superlinear scaling, or ``densification power law''~\cite{Leskovec2007CA_Full,Bettencourt2009}, i.e., $M$ grows in $N$ more than proportionally. This behavior is also evident in other days which we analyzed for each data set (Figs.~\ref{fig:cs1np_p1}--\ref{fig:cs1k_p1}). 
However, the scaling pattern emerges as a mixture of two distinct behaviors~\cite{kobayashi2020densification,kobayashi2021switching}; the straight-line scaling pattern indicates a constant exponent $\gamma > 1$ of $M\propto N^\gamma$, and it emerges for small to intermediate values of $N$. However, for larger values of $N$, total edges $M$ grows along an upward bending trajectory, implying an accelerating growth of $M$ in $N$. The two patterns are easily distinguished in the conference networks but to a lesser extent in Hospital and Workplace data.

In all data sets, a linear pattern tends to emerge within a specific range of values for population size $\Npht$ and activity level $\kapht$. 
Population size gradually expands from small to moderate values and, along with this,  $N_t$ is also increasing (Fig.~\ref{fig:csnp_p1}). At the same time, activity level is high in small networks with the number of edges $M$ at its upper bound $N(N-1)/2$ in some instances, implying that a considerable proportion of the individuals present are engaged (Figs.~\ref{fig:csk_p1}a and \ref{fig:cs1k_p1}a, c--d). 
However, as the population grows, activity level declines rapidly and $M$ continues to grow at a constant rate (e.g. Fig.~\ref{fig:csk_p1}a blue-green-yellow transition). During this phase, therefore, the dynamics between $M$ and $N$ are dominated by the gradual expansion of $\Npht$ which allows more and more individuals to become active. 

Given that population size of face-to-face networks is finite, $\Npht$ will eventually become constant but $M$ may continue to grow as the number of active nodes $N$ gradually approaches $\Npht$, yielding an upward bending slope towards $M$’s upper bound $N(N-1)/2$ (dotted line in Figs.~\ref{fig:csnp_p1} and \ref{fig:csk_p1}). The plots for IC2S2-17 and WS-16 in Figs.~\ref{fig:csnp_p1} and \ref{fig:csk_p1} suggest that this accelerating growth in $M$ occurs as $\hat{\kappa}$ increases while $\Npht$ remains high and relatively constant.
As the number of active individuals $N$ gets closer to $\Npht$, few isolated nodes (if any) remain, thus resulting in denser networks in which $M$ is almost at the maximum number of edges that can exist between active nodes. To enable these previously isolated individuals to make at least one connection, overall activity level increases, and this drives the continued growth in aggregate edges.
We also show in Supplementary Information the corresponding figures based on the alternative probability of connection in Eq~\eqref{eq:probfn0}, and the results are consistent with that of the baseline model 
(Figs.~\ref{fig:csnp_p0}--\ref{fig:cs1k_p0}).


\section*{Discussion}

In this study, we proposed a method to identify the driving force of the dynamical relationship between total active nodes $N$ and total edges $M$ in temporal networks. Changes in population size $\Np$ and overall activity $\kappa$ have both been identified as the mechanisms behind this dynamical relationship, each contributing to the emergence of different densification scaling patterns. Our main contribution is a numerical maximum-likelihood method that is able to estimate simultaneously, population size $\Np$ and activity rhythm $\kappa$ at given times, extending previous works in which one parameter is estimated by assuming the other is constant~\cite{kobayashi2020densification,kobayashi2021switching}. 
We found that changes in the mechanisms of densification and sparsification reflect explicit periodic transitions in networks that have rigid time constraints. 
Furthermore, our findings remain consistent with previous studies which explain the emergence of a constant scaling exponent as the result of an increasing population size, while the accelerated growth pattern is being impelled by intensification of overall activity. 

Although we have focused on social temporal networks in face-to-face contexts, the method is adaptable to practically any dynamical system that can be modeled as a time-varying network of nodes and edges. This is one advantage of our method because of the accessibility of $N$ and $M$ in most networks without having privacy issues. 
Of course, there are some limitations which need to be addressed in future research. Firstly, we employed a dynamic hidden variable model in generating networks, in which each node is randomly linked to another based on their individual activity. This means that although the model can reproduce the global quantities of $N$ and $M$, more realistic structural features that are known to exist in social networks (e.g. community structure, triads) are absent in generated networks. However, our focus in this work is to understand the variation in these global quantities of networks which does not require knowledge of structural properties. Our method also facilitates the use of network generating models that incorporate such properties observed in empirical networks.

Secondly, we assume that the distribution of node fitness (i.e., intrinsic activity of a node) in the network generating model is uniform. Although an empirical fitness distribution is preferred, the challenge exists in obtaining the individual activity level of nodes that are part of the population but are dormant (i.e., having no edges). Such nodes are generally not observable, because they are not explicitly stated as nodes that have interacted with others in the contact data set.

The relevance of this work lies in the simplicity of the method for understanding the dynamical relationship between fundamental global quantities of temporal networks, and the adaptability of our method to include more realistic features of empirical networks. The dynamics of network growth and shrinkage is central to how systems work, and it would also be one crucial factor in how information and infectious diseases  spread in networks. Given the pervasiveness of complex systems and our reliance on them in our daily lives, greater understanding of the dynamics of networks would improve how we interact with, and even control such systems.

\section*{Acknowledgments}
T.\ K. acknowledges financial support from JSPS KAKENHI\ 19H01506 and 20H05633.

\section*{Data availability}
The data and Python code are available in GitHub~\cite{SF_github}.



\clearpage
\setcounter{section}{0}
\setcounter{table}{0}
\setcounter{equation}{0}
\setcounter{figure}{0}
\setcounter{page}{1}
     
\renewcommand{\thetable}{S\arabic{table}}
\renewcommand{\thefigure}{S\arabic{figure}}
\renewcommand{\thesection}{S\arabic{section}}
\renewcommand{\theequation}{S\arabic{equation}}

{\flushleft
{\fontsize{16pt}{16pt}\selectfont
 \textbf{Supplementary Information:} \\
 \vspace{.7cm}
 \Large{``Identifying the temporal dynamics of densification and sparsification in human contact networks"} \\
 \vspace{.5cm}
{\large Shaunette Ferguson and Teruyoshi Kobayashi}
}
}
\vspace{2cm}

\begin{figure}[ht]
    \centering
    \includegraphics[width = 0.85\textwidth]{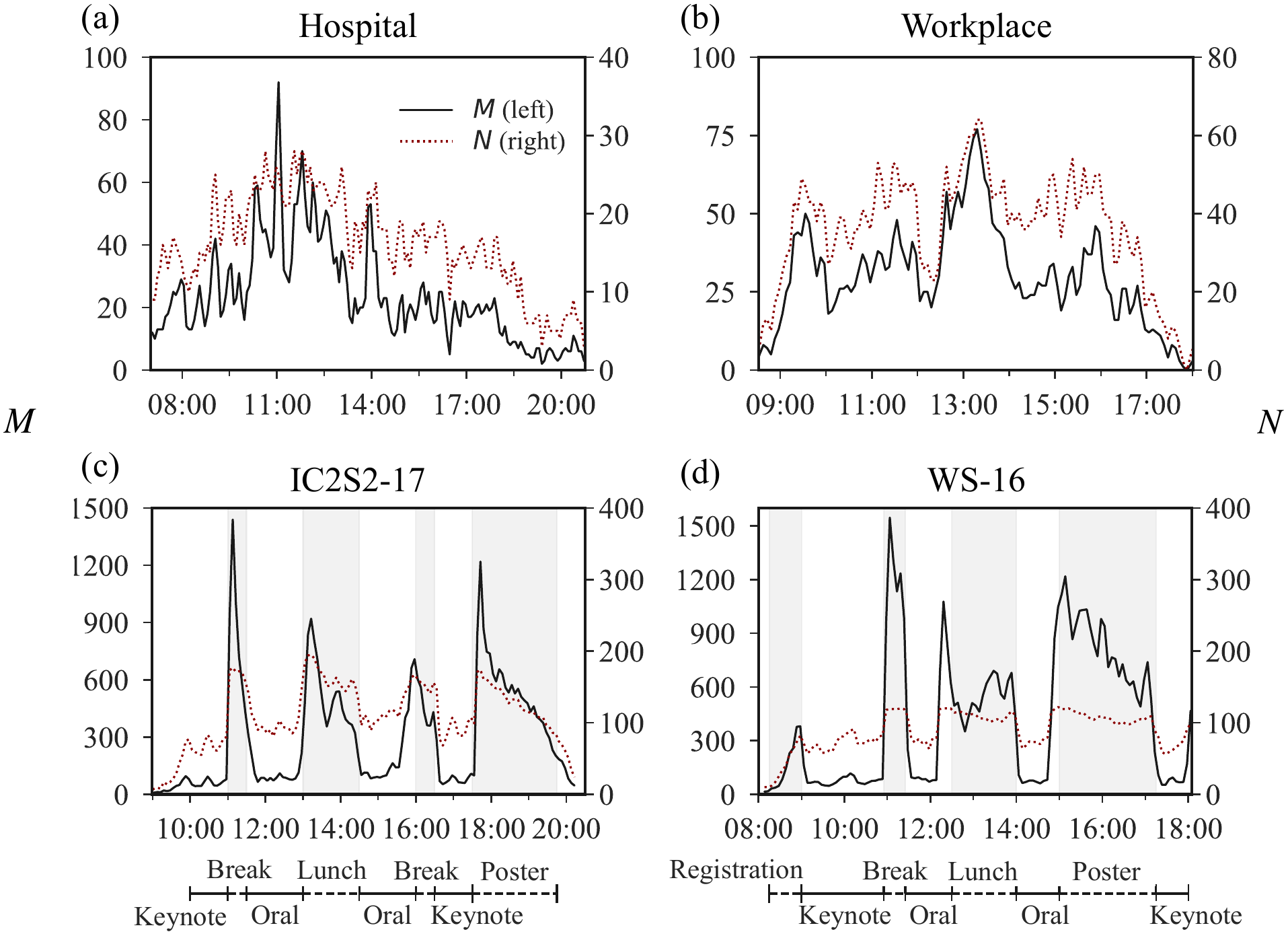}
    \caption{Evolution of total active edges $M$ and total active nodes $N$ in face-to-face networks. The following days are shown for each data set: (a) Hospital on December 8, 2010, (b) Workplace on June 28, 2015, (c) IC2S2-17 on July 12, 2017 and (d) WS-16 on November 30, 2016. Timeline below conference data ((c) IC2S2-17 and (d) WS-16) identify time windows for scheduled events. Gray shading highlights unrestricted sessions i.e., registration, break, lunch and poster session.}
    \label{fig:MNline1}
\end{figure}

\clearpage

\begin{figure}[t]
    \centering
    \includegraphics[scale=0.93]{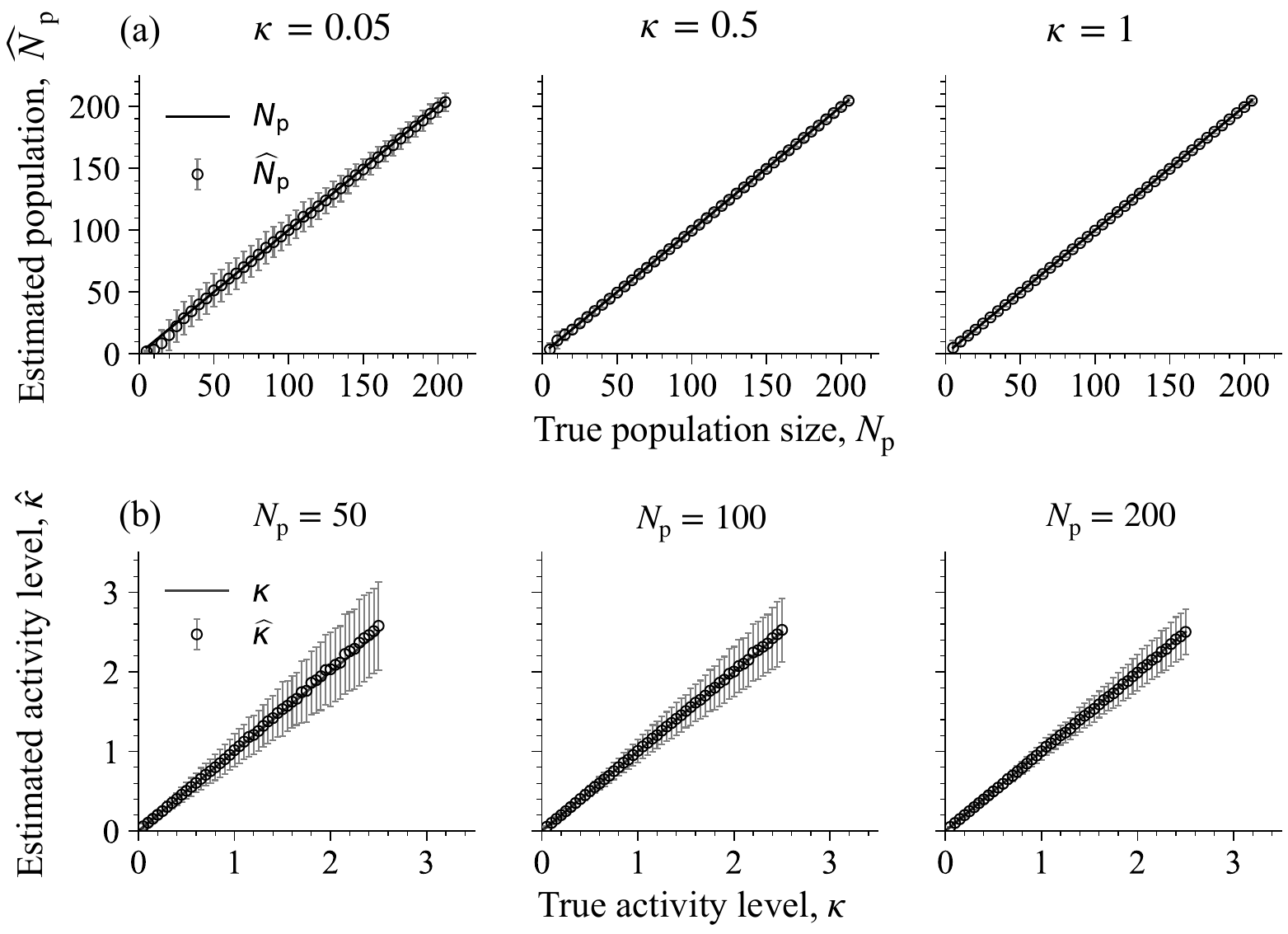}
    \caption{Validation of the maximum-likelihood estimation method. (a) Estimation of $\Np$ for $\Np$ ranging from 5 to 200 (incremented by 5) given $\kappa = 0.05$, $\kappa = 0.5$  and $\kappa = 1$. (b) Estimation of $\kappa$ for $\kappa$ ranging from 0.05 to 2.5 (incremented by 0.05) given $\Np = 50$, $\Np = 100$  and $\Np = 200$. The estimates $\Npht$ and $\kapht$ are obtained from the model with probability of a connection between two nodes $i$ and $j$ given by Eq.~\eqref{eq:probfn}. Errors bars represent one standard deviation and they are computed over 1,000 runs.}
    \label{fig:validp1_alternating}
\end{figure}

\clearpage

\begin{figure}[tb]
    \centering
    \includegraphics[scale=0.85]{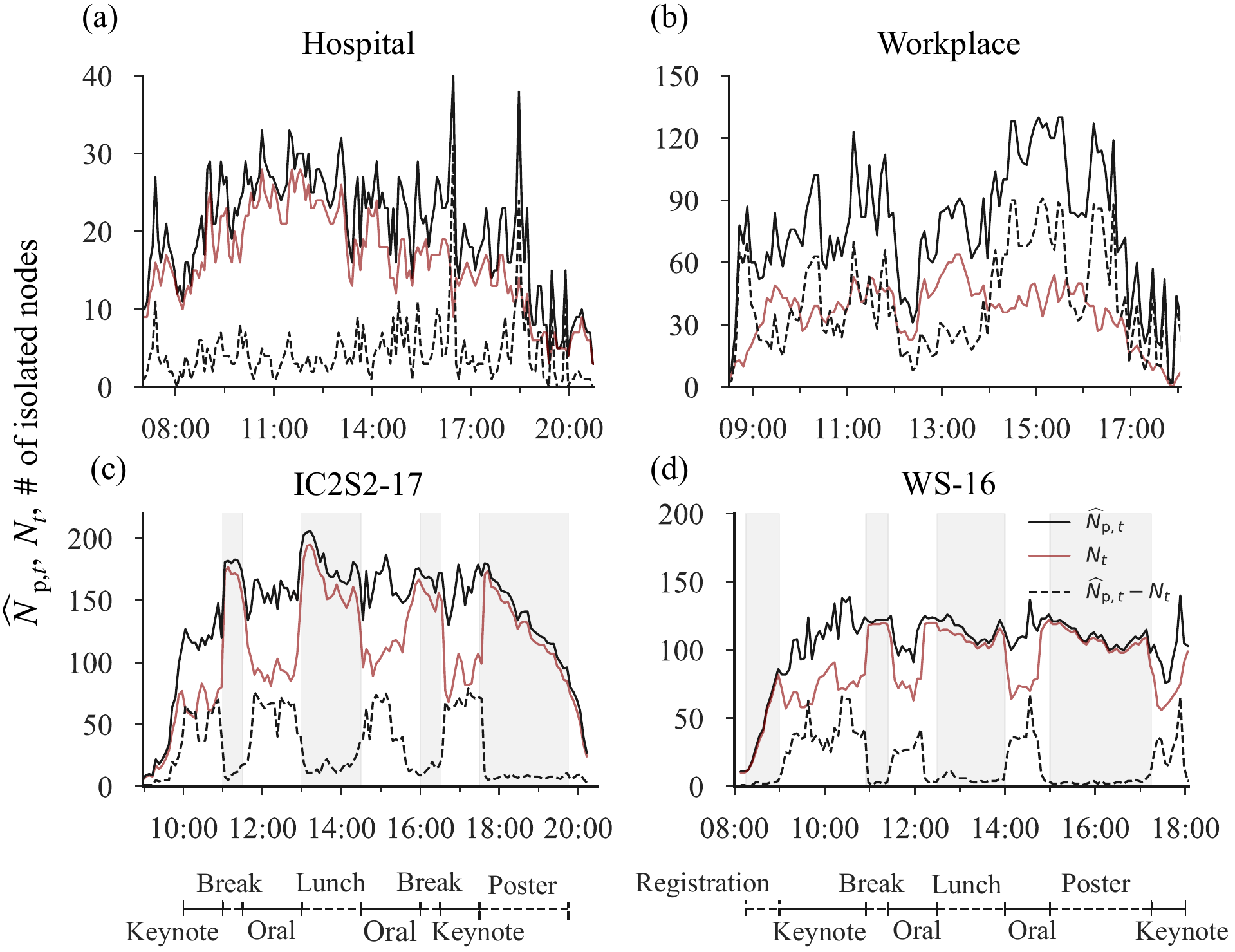}
    \caption{Shifts in estimated population size $\Npht$, number of active persons $N_t$ and total isolated nodes $\Npht -N_t$ in face-to-face networks. The following days are shown for each data set: (a) Hospital on December 8, 2010 (b) Workplace on June 28, 2015 (c) IC2S2-17 on July 12, 2017  and (d) WS-16 on November 30, 2016. Timelines below conference data, in panels (c) and (d), identify time windows for scheduled events. Gray shading highlights unrestricted sessions i.e., registration, break, lunch and poster session. Estimates of $\Npt$ are based on the model with probability $p_{ij,t} = 1-e^{-\kappa a_ia_j}$ from Eq.~\eqref{eq:probfn}.}
    \label{fig:p1estimates_np1}
\end{figure}

\begin{figure}[t!]
    \centering
    \includegraphics[scale=0.85]{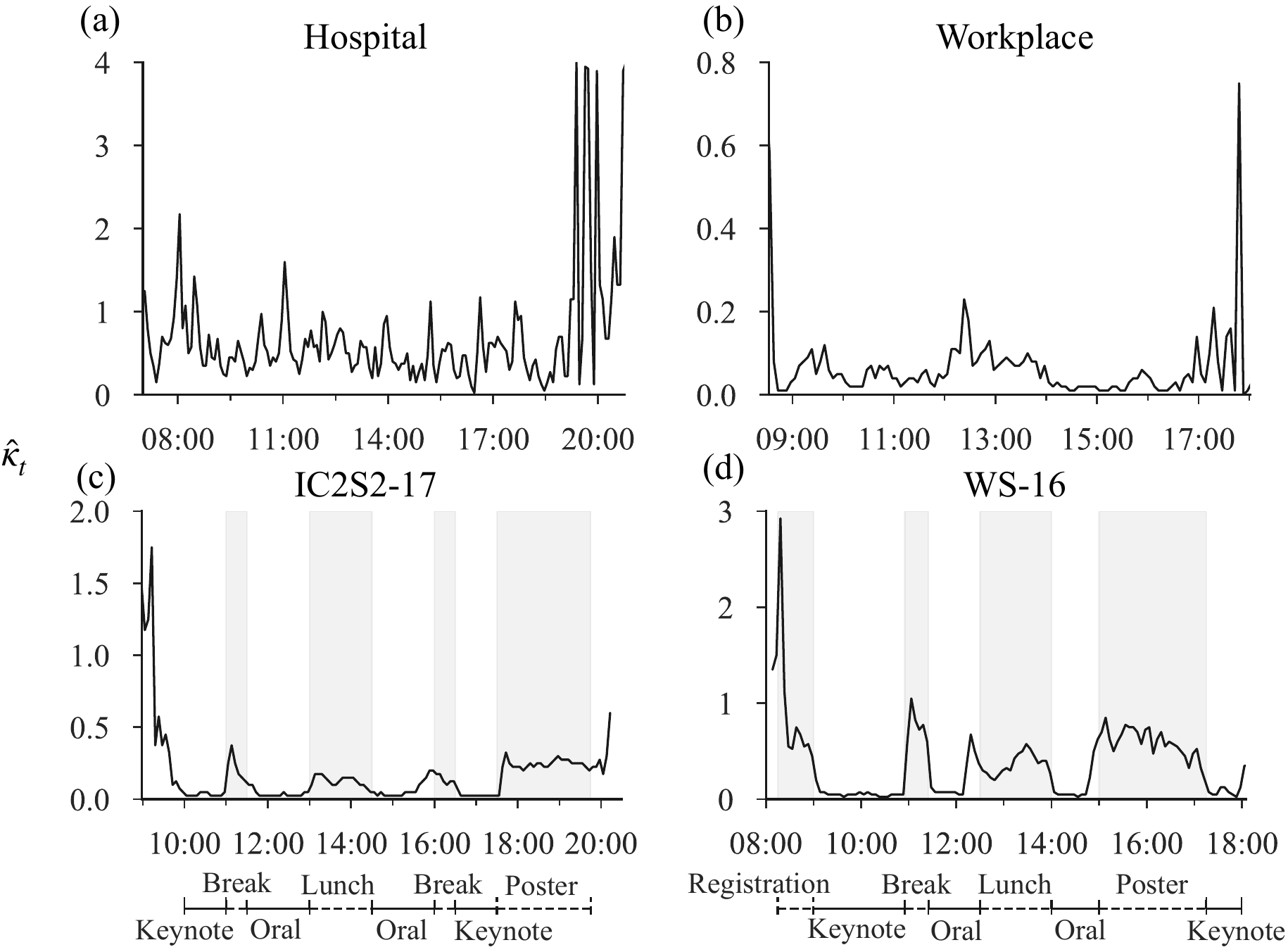}
    \caption{Changes in estimated overall activity $\kapht$ in face-to-face networks. The following days are shown for each data set: (a) Hospital on December 8, 2010 (b) Workplace on June 28, 2015 (c) IC2S2-17 on July 12, 2017  and (d) WS-16 on November 30, 2016. Timelines below conference data (c) IC2S2-17 and (d) WS-16 identify time windows for scheduled events. Gray shading highlights unrestricted sessions i.e., registration, break, lunch and poster session. Estimates of $\kapht$ are based on the model with probability $p_{ij,t} = 1-e^{-\kappa a_ia_j}$ from Eq.~\eqref{eq:probfn}.}
    \label{fig:p1estimates_k1}
\end{figure}

\clearpage

\begin{figure}[tb]
    \centering
    \includegraphics[scale=0.85]{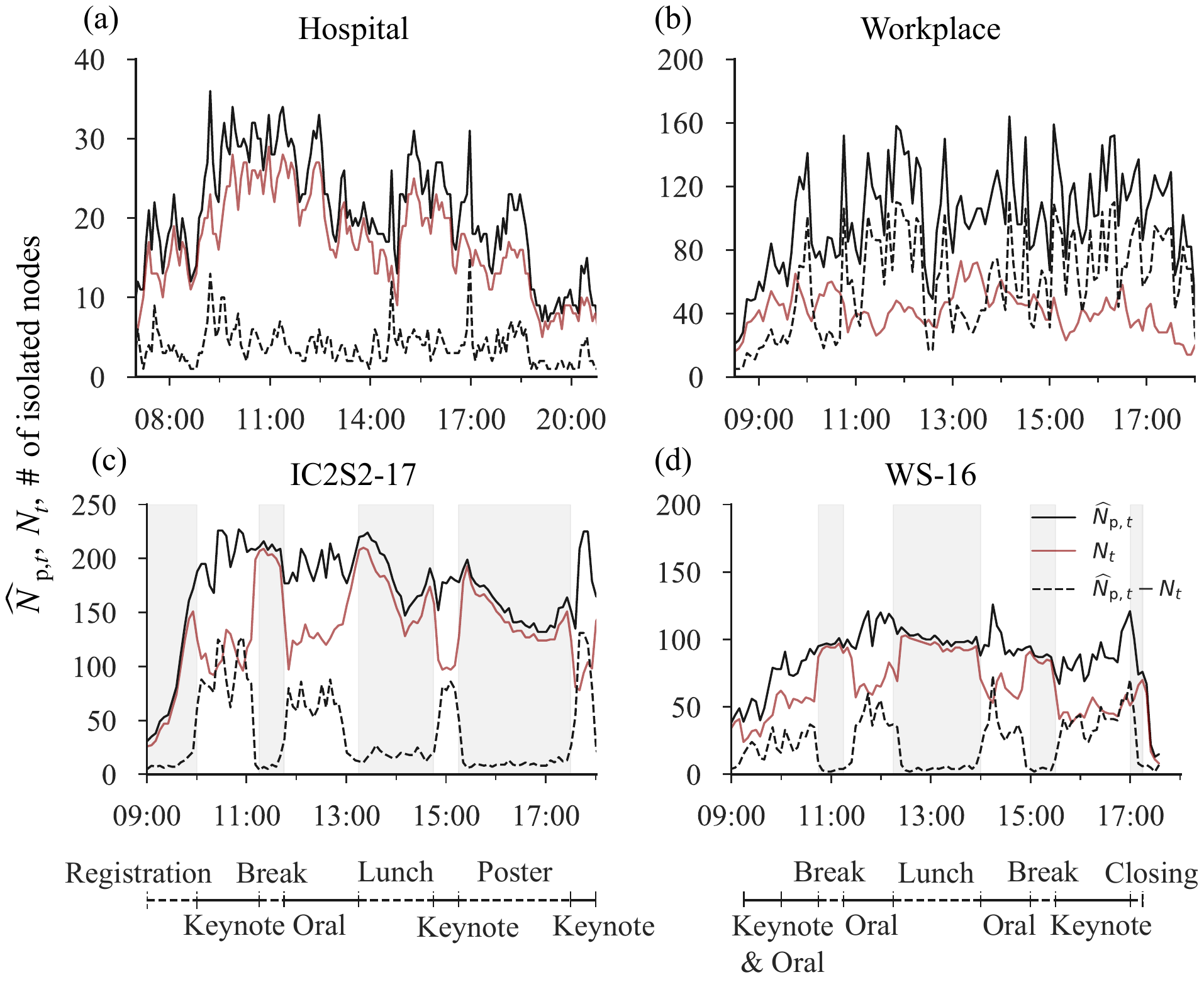}
    \caption{Shifts in estimated population size $\Npht$, number of active persons $N_t$ and total isolated nodes $\Npht -N_t$ in face-to-face networks. The following days are shown for each data set: (a) Hospital on December 7, 2010 (b) Workplace on June 27, 2015 (c) IC2S2-17 on July 11, 2017 and (d) WS-16 on December 1, 2016. Timelines below conference data (c) IC2S2-17 and (d) WS-16 identify time windows for scheduled events. Gray shading highlights unrestricted sessions i.e., registration, break, lunch, poster session and closing remarks. Estimates of $\Npt$ are based on the model with probability $p_{ij,t} = \kappa a_ia_j$ from Eq.~\eqref{eq:probfn0}.}
    \label{fig:p0estimates_np}
\end{figure}

\begin{figure}[t!]
    \centering
    \includegraphics[scale=0.85]{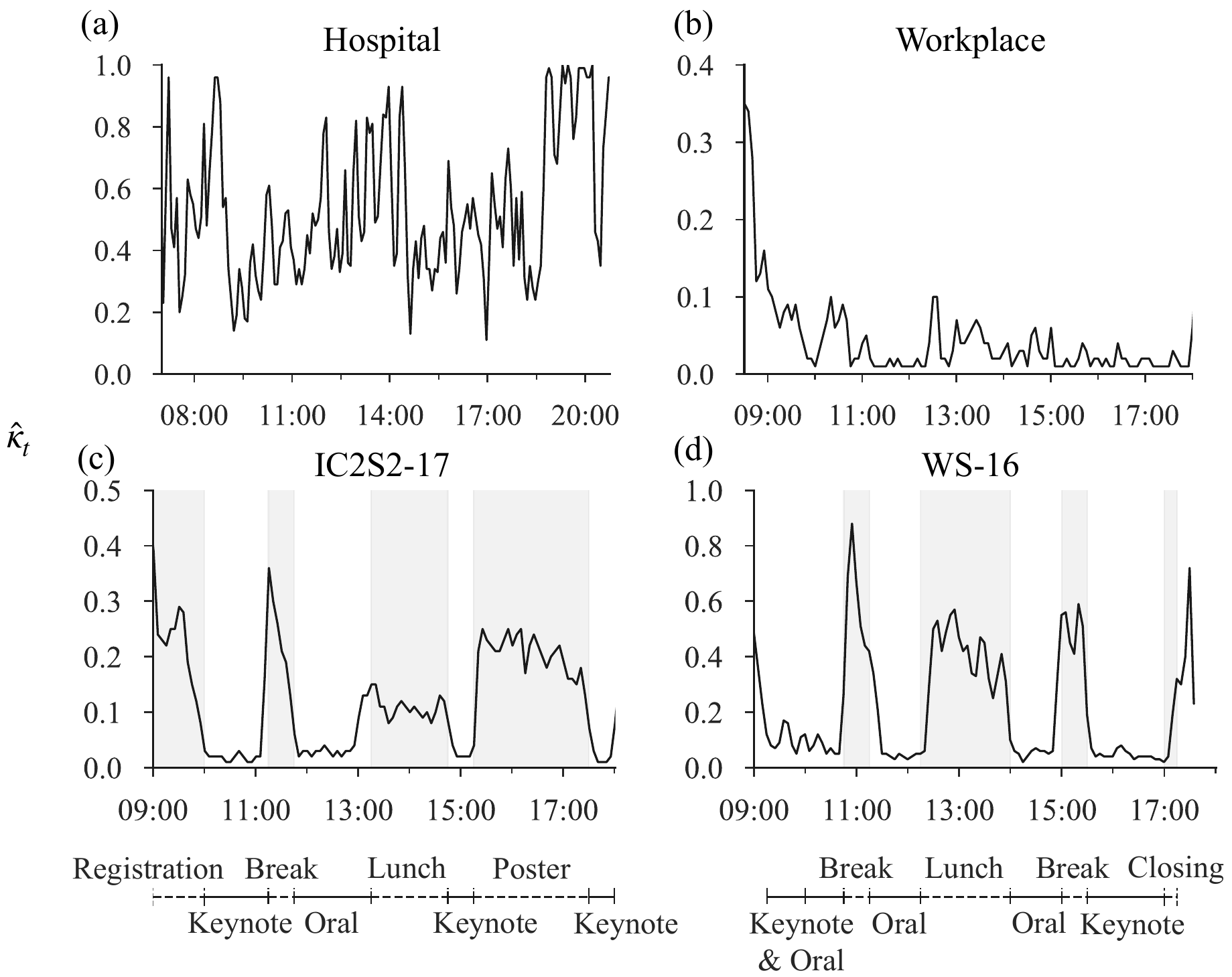}
    \caption{Changes in estimated overall activity $\kapht$ in face-to-face networks. The following days are shown for each data set: (a) Hospital on December 7, 2010 (b) Workplace on June 27, 2015 (c) IC2S2-17 on July 11, 2017 and (d) WS-16 on December 1, 2016. Timelines below conference data (c) IC2S2-17 and (d) WS-16 identify time windows for scheduled events. Gray shading highlights unrestricted sessions i.e., registration, break, lunch, poster session and closing remarks. Estimates of $\kapht$ are based on the model with probability $p_{ij,t} = \kappa a_ia_j$ from Eq.~\eqref{eq:probfn0}.}
    \label{fig:p0estimates_kp}
\end{figure}

\begin{figure}[tb]
    \centering
    \includegraphics[scale=0.85]{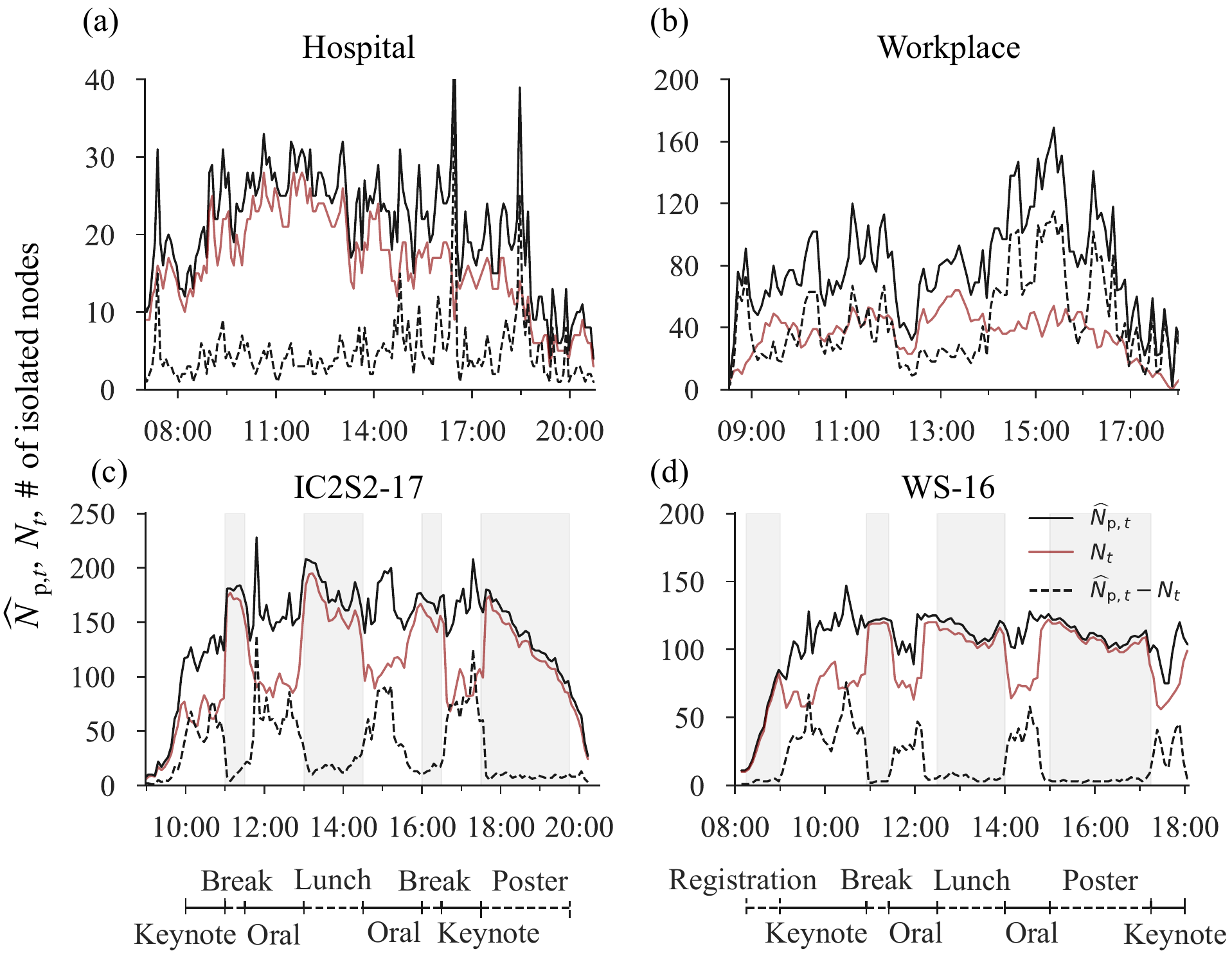}
    \caption{Shifts in estimated size of population $\Npht$, number of active persons $N_t$ and total isolated nodes $\Npht -N_t$ in face-to-face networks. The following days are shown for each data set: (a) Hospital on December 8, 2010 (b) Workplace on June 28, 2015 (c) IC2S2-17 on July 12, 2017 and (d) WS-16 on November 30, 2016. Timelines below conference data (c) IC2S2-17 and (d) WS-16 identify time windows for scheduled events. Gray shading highlights unrestricted sessions i.e., registration, break, lunch and poster session. Estimates of $\Npt$ are based on the model with probability $p_{ij,t} = \kappa a_ia_j$ from Eq.~\eqref{eq:probfn0}.}
    \label{fig:p0estimates_np1}
\end{figure}

\clearpage

\begin{figure}[t!]
    \centering
    \includegraphics[scale=0.85]{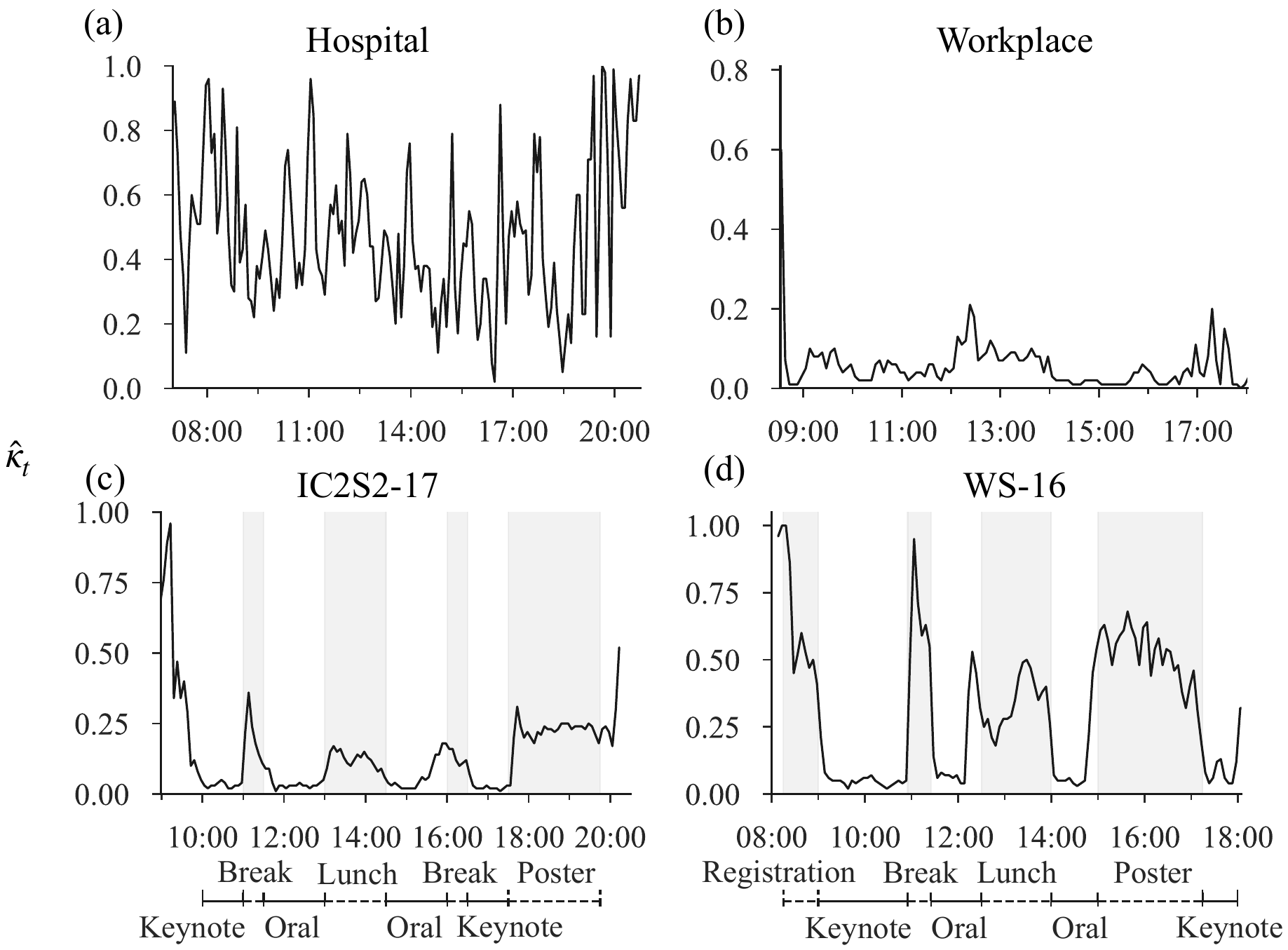}
    \caption{Changes in estimated overall activity $\kapht$ in face-to-face networks. The following days are shown for each data set: (a) Hospital on December 8, 2010 (b) Workplace on June 28, 2015 (c) IC2S2-17 on July 12, 2017 and (d) WS-16 on November 30, 2016. Timelines below conference data (c) IC2S2-17 and (d) WS-16 identify time windows for scheduled events. Gray shading highlights unrestricted sessions i.e., registration, break, lunch, and poster session. Estimates of $\kapht$ are based on the model with probability $p_{ij,t} = \kappa a_ia_j$ from Eq.~\eqref{eq:probfn0}.}
    \label{fig:p0estimates_k1}
\end{figure}

\begin{figure}[t]
    \centering
    \includegraphics[scale = 0.85]{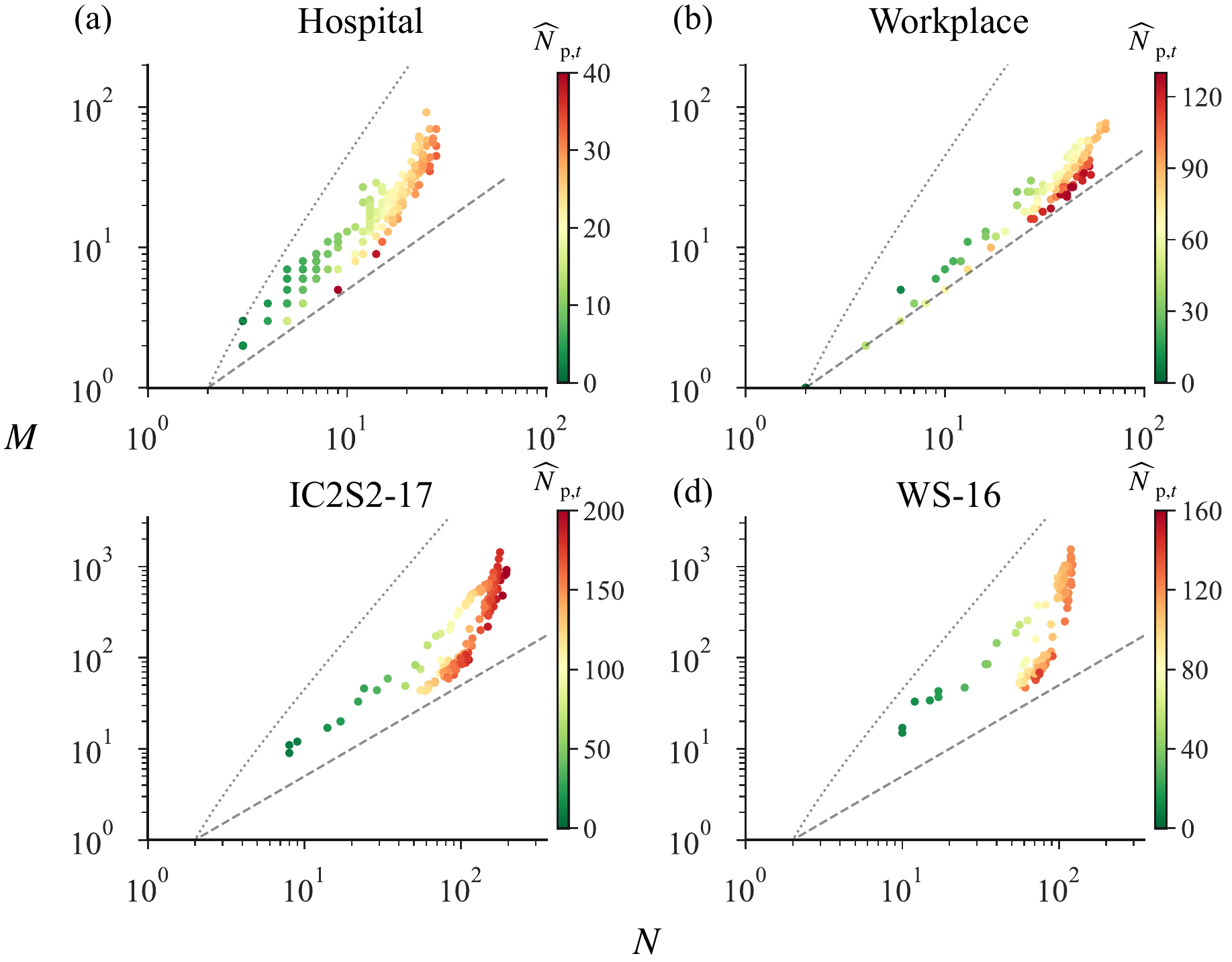}
    \caption{Densification scaling and changes in estimated population size in face-to-face networks. $N$-$M$ scaling plots are shown for: (a) Hospital on December 8, 2010 (b) Workplace on June 28, 2015 (c) IC2S2-17 on July 12, 2017 and (d) WS-16 on November 30, 2016. Each marker represents a snapshot of the network and colors denote estimated population size $\Npht$ based on the respective color bar. Gray dashed and dotted lines show theoretical lower ($M = N/2$) and upper ($M = N(N-1)/2$) bounds. Estimates of $\Npt$ based on the model with probability $p_{ij,t} = 1-e^{-\kappa a_ia_j}$ from Eq.~\eqref{eq:probfn}.}
    \label{fig:cs1np_p1}
\end{figure}

\begin{figure}[t]
    \centering
    \includegraphics[scale = 0.85]{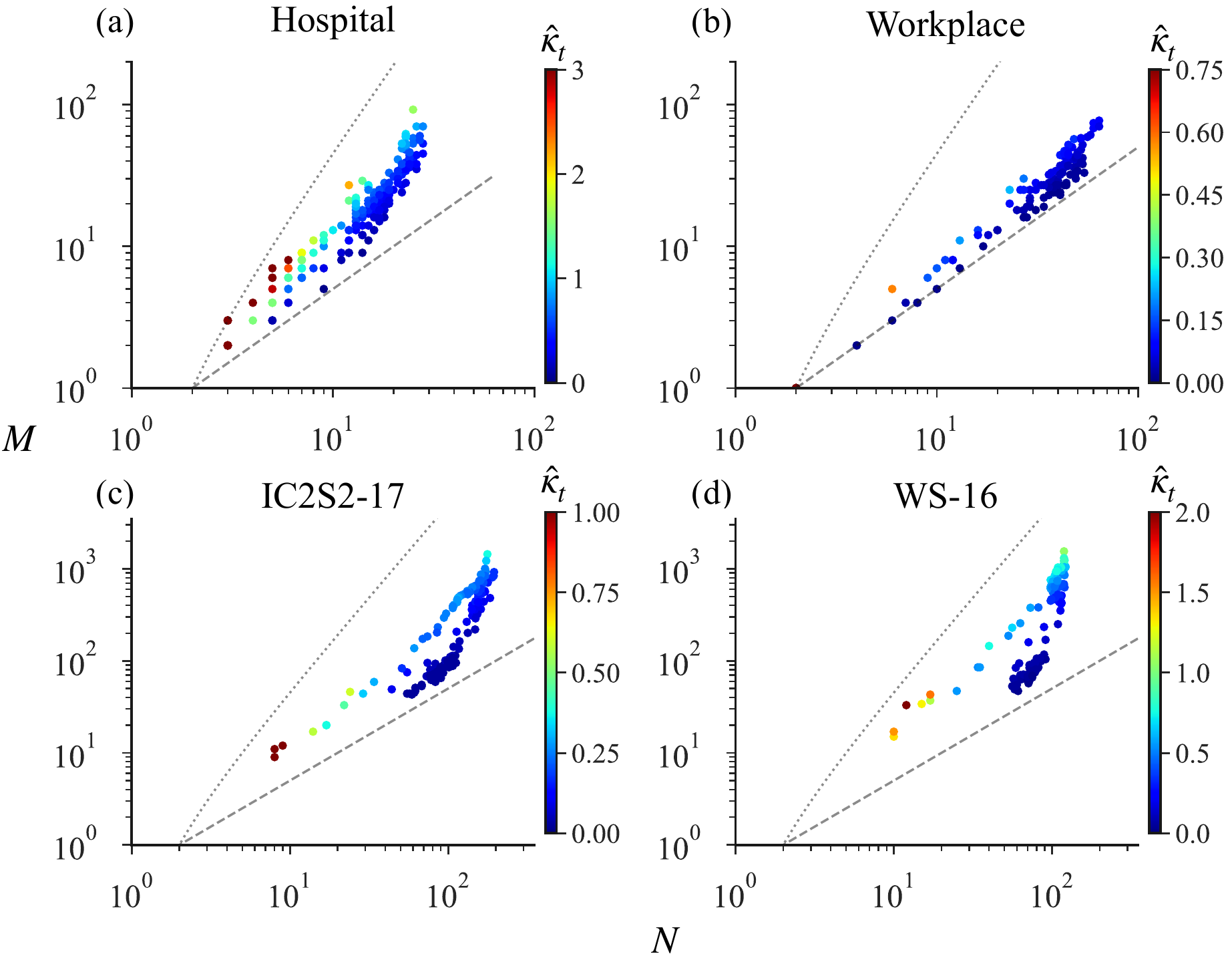}
    \caption{Densification scaling and changes in estimated overall activity in face-to-face networks. $N$-$M$ scaling plots are shown for: (a) Hospital on December 8, 2010 (b) Workplace on June 28, 2015 (c) IC2S2-17 on July 12, 2017 and (d) WS-16 on November 30, 2016. Each marker represents a snapshot of the network and colors denote estimated activity level $\kapht$ based on the respective color bar. Gray dashed and dotted lines show theoretical lower ($M = N/2$) and upper ($M = N(N-1)/2$) bounds. Estimates of $\kapht$ based on the model with probability $p_{ij,t} = 1-e^{-\kappa a_ia_j}$ from Eq.~\eqref{eq:probfn}.}
    \label{fig:cs1k_p1}
\end{figure}
 
\clearpage
\clearpage

\begin{figure}[t]
    \centering
    \includegraphics[scale = 0.85]{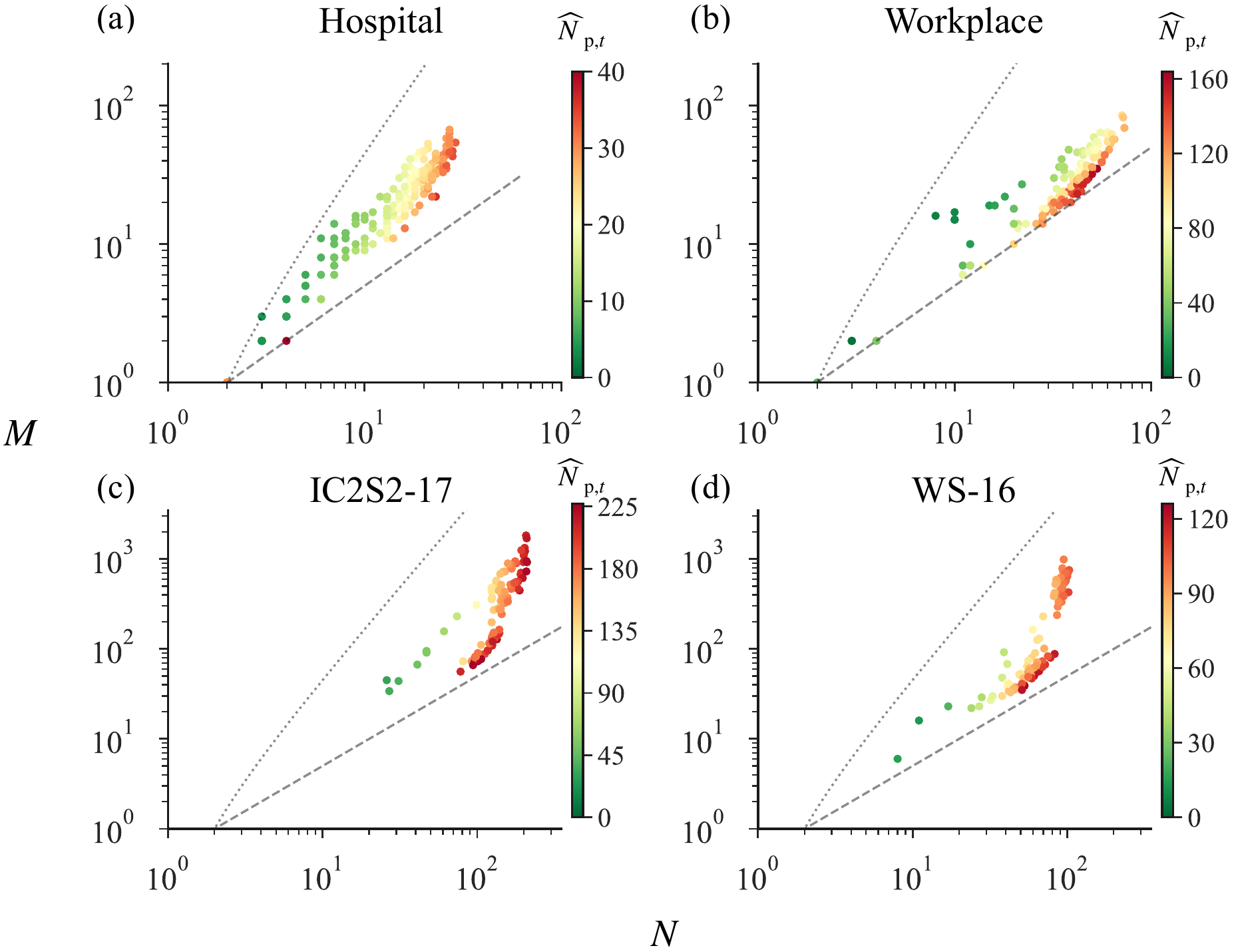}
    \caption{Densification scaling and changes in estimated population size in face-to-face networks. $N$-$M$ scaling plots are shown for: a) Hospital on December 7, 2010 (b) Workplace on June 27, 2015 (c) IC2S2-17 on July 11, 2017 and (d) WS-16 on December 1, 2016. Each marker represents a snapshot of the network and colors denote estimated population size $\Npht$ based on the respective color bar. Gray dashed and dotted lines show theoretical lower ($M = N/2$) and upper ($M = N(N-1)/2$) bounds. Estimates of $\Npt$ based on the model with probability $p_{ij,t} = \kappa a_ia_j$ from Eq.~\eqref{eq:probfn0}.}
    \label{fig:csnp_p0}
\end{figure}

\clearpage

\begin{figure}[t]
    \centering
    \includegraphics[scale = 0.85]{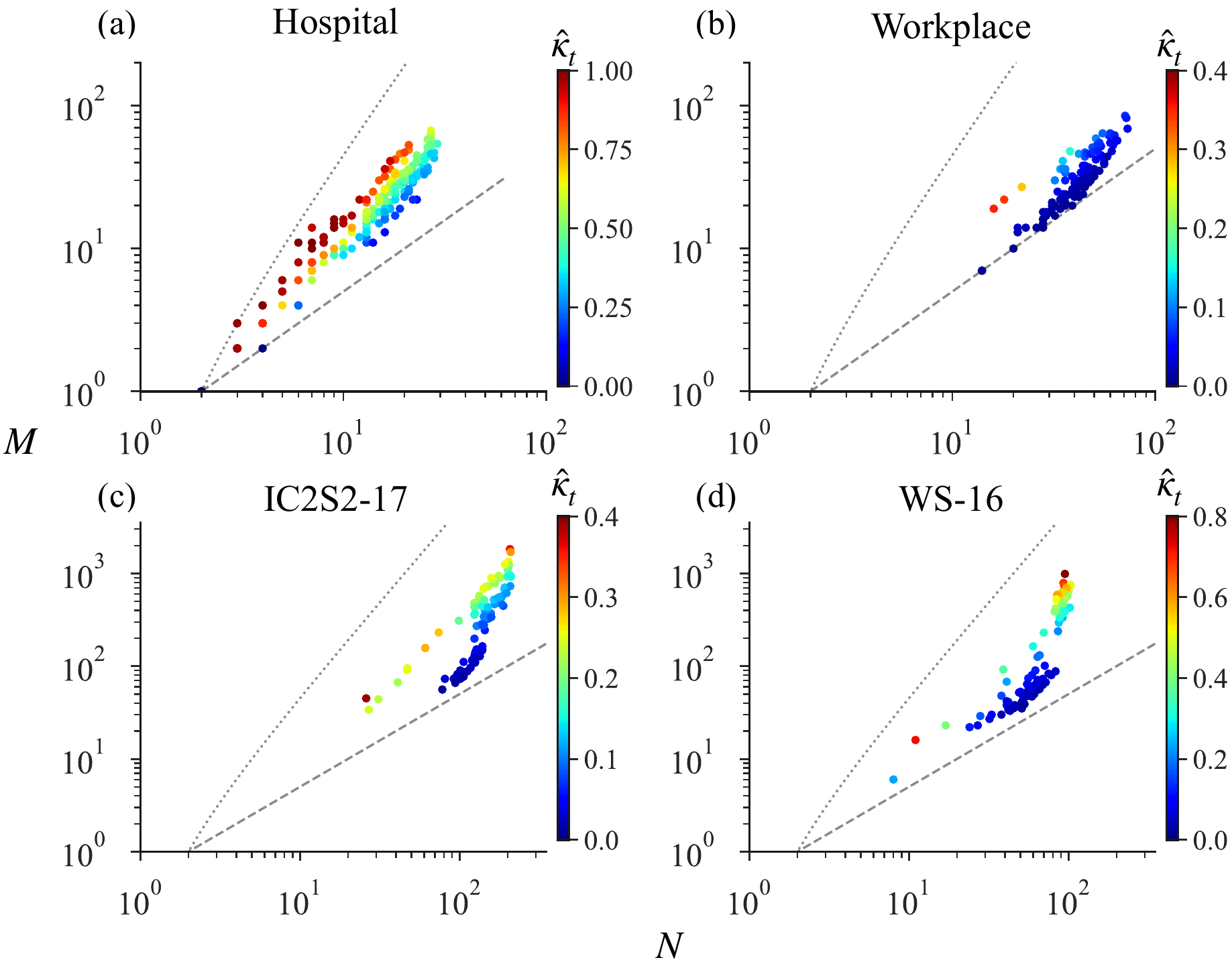}
    \caption{Densification scaling and changes in estimated overall activity in face-to-face networks. $N$-$M$ scaling plots for: a) Hospital on December 7, 2010 (b) Workplace on June 27, 2015 (c) IC2S2-17 on July 11, 2017 and (d) WS-16 on December 1, 2016. Each marker represents a snapshot of the network and colors denote estimated activity level $\kapht$ based on the respective color bar. Gray dashed and dotted lines show theoretical lower ($M = N/2$) and upper ($M = N(N-1)/2$) bounds. Estimates of $\kapht$ based on the model with probability $p_{ij,t} = \kappa a_ia_j$ from Eq.~\eqref{eq:probfn0}.}
    \label{fig:csk_p0}
\end{figure}

\clearpage

\clearpage

\begin{figure}[t]
    \centering
    \includegraphics[scale = 0.85]{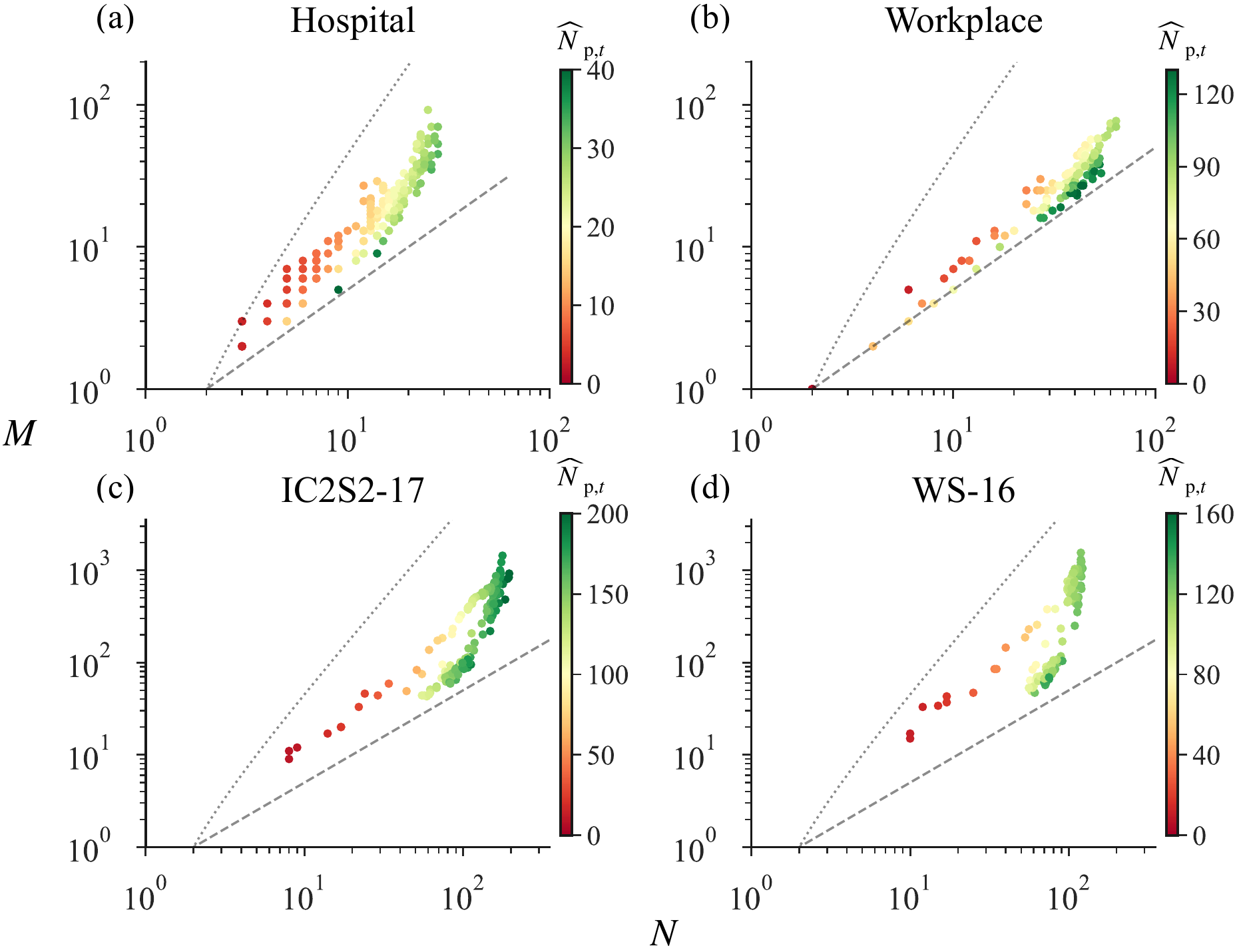}
    \caption{Densification and changes in estimated population size in face-to-face networks. $N$-$M$ scaling plots are shown for: (a) Hospital on December 8, 2010 (b) Workplace on June 28, 2015 (c) IC2S2-17 on July 12, 2017 and (d) WS-16 on November 30, 2016. Each marker represents a snapshot of the network and colors denote estimated population size $\Npht$ based on the respective color bar. Gray dashed and dotted lines show theoretical lower ($M = N/2$) and upper ($M = N(N-1)/2$) bounds. Estimates of $\Npt$ based on the model with probability $p_{ij,t} = \kappa a_ia_j$ from Eq.~\eqref{eq:probfn0}.}
    \label{fig:cs1np_p0}
\end{figure}

\clearpage

\begin{figure}[t]
    \centering
    \includegraphics[scale = 0.85]{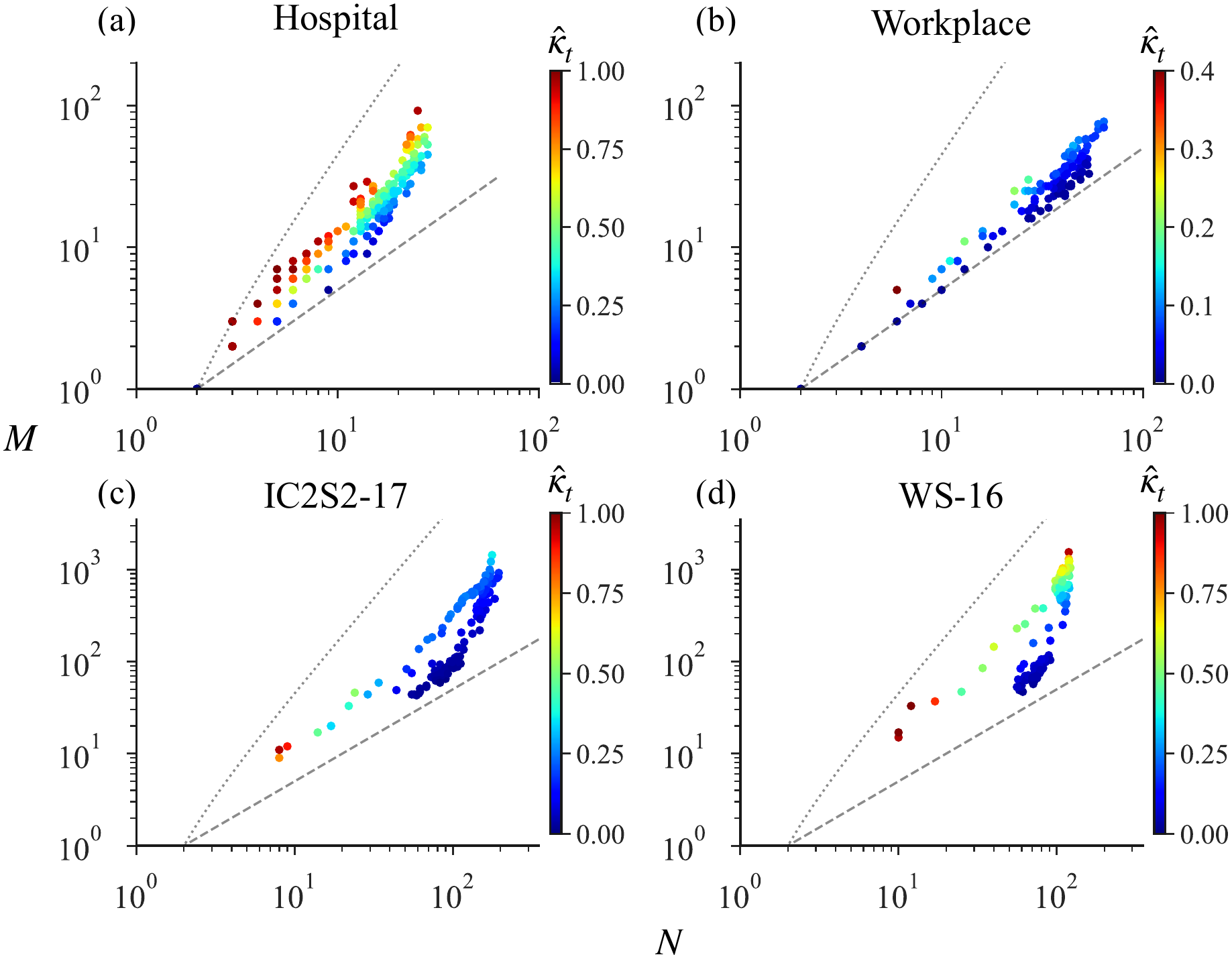}
    \caption{Densification and changes in the estimated overall activity in face-to-face networks. $N$-$M$ scaling plots are shown for: (a) Hospital on December 8, 2010 (b) Workplace on June 28, 2015 (c) IC2S2-17 on July 12, 2017 and (d) WS-16 on November 30, 2016. Each marker represents a snapshot of the network and colors denote estimated activity level $\kapht$ based on the respective color bar. Gray dashed and dotted lines show theoretical lower ($M = N/2$) and upper ($M = N(N-1)/2$) bounds. Estimates of $\kapht$ based on the model with probability $p_{ij,t} = \kappa a_ia_j$ from Eq.~\eqref{eq:probfn0}.}
    \label{fig:cs1k_p0}
\end{figure}





\end{document}